\documentclass[english,notitlepage,a4paper,nofootinbib]{revtex4-1}
\usepackage[utf8]{inputenc}
\usepackage{babel}
\usepackage{textcomp}
\usepackage{amsmath}
\usepackage{graphicx}
\usepackage{wasysym}
\usepackage[normalem]{ulem}

\long\def\beginpgfgraphicnamed#1#2\endpgfgraphicnamed{\includegraphics{#1}}

\usepackage{xcolor}
\definecolor{lcolor}{rgb}{0.5,0,0}
\definecolor{citcolor}{rgb}{0,0.3,0.0}
\usepackage[unicode=true,pdfusetitle,bookmarks=true,bookmarksnumbered=false,bookmarksopen=false,breaklinks,colorlinks,urlcolor=blue,citecolor=citcolor,linkcolor=lcolor]{hyperref}

\newcommand{\qs}{Q_\mathrm{s}}
\newcommand{\as}{\alpha_\mathrm{s}}
\newcommand{\nr}[1]{(\ref{#1})} 
\newcommand{\fig}{Fig.~}

\newcommand{\eq}{Eq.~}

\newcommand{\eqs}{Eqs.~}

\makeatletter
\usepackage{amsmath}
\usepackage{slashed}
\usepackage[compat=1.1.0]{tikz-feynman}

\makeatother

\begin{document}

\title{The dipole picture and the non-relativistic expansion}

\author{Miguel \'{A}ngel Escobedo}

\affiliation{Department of Physics, P.O. Box 35, FI-40014 University of Jyv\"{a}skyl\"{a},
Finland}

\affiliation{Instituto Galego de F\'{i}sica de Altas Enerxías (IGFAE), Universidade
de Santiago de Compostela. E-15782, Galicia, Spain}

\author{Tuomas Lappi}
\affiliation{Department of Physics, P.O. Box 35, FI-40014 University of Jyv\"{a}skyl\"{a},
Finland}
\affiliation{
Helsinki Institute of Physics, P.O. Box 64, 00014 University of Helsinki,
Finland
}

\begin{abstract}
We study exclusive quarkonium production in the dipole picture at next-to-leading order (NLO) accuracy, using the non-relativistic expansion for the quarkonium wavefunction. This process offers one of the best ways to obtain information about gluon distributions at small $x$, in ultraperipheral heavy ion collisions and in deep inelastic scattering.  The quarkonium light cone wave functions needed in the dipole picture have typically been available only at tree level, either in phenomenological models or in the nonrelativistic limit. In this paper, we discuss the compatibility of the dipole approach and the non-relativistic expansion and  compute NLO
 relativistic corrections to the quarkonium light-cone wave function
in light-cone gauge. Using these corrections we recover results for
the NLO decay width of quarkonium to $e^{+}e^{-}$ and we check that
the non-relativistic expansion is consistent with ERBL evolution and
with B-JIMWLK evolution of the target. The results presented here
will allow computing the exclusive quarkonium production rate at NLO
once the one loop photon wave function with massive quarks, currently
under investigation, is known.
\end{abstract}

\maketitle

\section{Introduction}

The partonic  structure of hadrons and nuclei in the limit of high collision energies, or equivalently small momentum fractions $x$, is poorly constrained by existing experimental data. It is believed that at high enough energies the properties of small-$x$ gluons are dominated by gluon saturation, i.e. the dominance of nonlinear interactions in the gluon field. In order to fully understand the behavior of small $x$ gluons, a variety of different experimental measurements is needed. Of particular importance here is exclusive quarkonium production mediated by real or virtual photons. Such measurements are currently made in ultraperipheral heavy ion collisions~\cite{Baltz:2007kq}  at the LHC and at RHIC. Exclusive measurements will also be an important part of the program  at a future electron-ion-collider~\cite{Accardi:2012qut}. Exclusive quarkonium production is an important process for several reasons. As an exclusive process it depends on the gluon density quadratically and is thus more sensitive to nonlinearities than inclusive cross sections. Exclusive processes can, depending on exactly what final state of the target one measures, be sensitive to separately the average and the fluctuations of the gluon density in the target~\cite{Caldwell:2009ke,Lappi:2010dd,Mantysaari:2016ykx,Mantysaari:2017dwh}. On the other hand, the heavy quark masses cut away nonperturbative long distance contributions and make the use of a weak coupling framework safer than for light quark processes~\cite{Mantysaari:2018nng,Mantysaari:2018zdd}.

The dipole picture of DIS~\cite{Nikolaev:1990ja,Nikolaev:1991et,Mueller:1993rr,Mueller:1994jq,Mueller:1994gb} 
(a specialization of the light cone perturbation theory framework of \cite{Bjorken:1970ah} to the DIS process) provides a convenient framework to study deep inelastic scattering at high energy.
In particular one expresses both inclusive and exclusive cross sections in terms of the same fundamental quantity, the dipole scattering amplitude, which gives this picture more predictive power than collinear factorization. With light quarks several recent advances have taken calculations of inclusive~\cite{Beuf:2016wdz,Beuf:2017bpd,Hanninen:2017ddy} and diffractive~\cite{Boussarie:2016ogo,Boussarie:2016bkq} observables to NLO accuracy, and it would be important to do the same for heavy quark cross sections. 

At leading order  the physical picture of exclusive scattering in the dipole picture is the following \cite{Brodsky:1994kf}: a virtual photon
fluctuates into a quark-antiquark pair that interacts with the nucleus
elastically with a cross-section $\sigma_{q\bar{q}}$. The
resulting dipole can, later on, recombine into a quarkonium state.
In this framework, the information about quarkonium is encoded in
its light-cone wave function (LCWF), which encodes the overlap of the quarkonium state with various eigenstates of the free light cone QCD Hamiltonian.

In recent phenomenological applications (following e.g.~\cite{Kowalski:2006hc}) one commonly uses phenomenological parametrizations for the LCWF, such as the the ``boosted gaussian'' or ``gaussLC''.
A more model-independent approach would be to exploit the fact that
heavy quarkonium is a non-relativistic system. In such a system the typical three-momentum of quarks is of the order of $mv$,where $m$ the heavy quark mass\footnote{In this paper, we use the pole mass in our computations. However, we note that in future phenomenological applications it might be convenient to translate the results to another scheme.} and $v$ the typical speed of the
heavy quarks around the center of mass (which is much smaller than
the speed of light). The mass of the heavy quarkonium state $M_{HQ}$
is such that the binding energy is small: $M_{HQ}-2m\sim mv^{2}\ll m$. Another important energy
scale is $\Lambda_{QCD}$ which marks the transition between perturbative
and non-perturbative physics in QCD.

Using an effective field theory approach non-relativistic QCD (NRQCD) \cite{Bodwin:1994jh}
factorization formulas for some processes can be proven. The NRQCD approach
has been quite successful, although there is some tension with polarization
related observables in charmonium (for a more recent review see \cite{Brambilla:2010cs}).
 Diffractive quarkonium production has been studied in the  non-relativistic limit in both a covariant \cite{Ryskin:1992ui} theory approach and in a light cone formalism \cite{Brodsky:1994kf}.
Also velocity expansion corrections have been discussed in \cite{Hoodbhoy:1996zg},
although they are a NNLO effect in $\alpha_{s}(mv)$. However, NLO corrections
in $\alpha_{s}(m)$ due to radiative corrections to the computation
in \cite{Ryskin:1992ui,Brodsky:1994kf} are still missing.

In this paper, we are going to compute the LO relativistic corrections
to quarkonium light-cone wave function in the light-cone gauge such
that it can be used to obtain NLO predictions in the dipole picture.
This will allow to compute the radiative NLO corrections to the leading nonrelativistic result. We are going to check that using our results
for the quarkonium light-cone wave function we can obtain the well-known
literature result of the NLO correction to quarkonium decay to $e^{+}e^{-}$
\cite{Barbieri:1975ki}. We will then show that the light-cone distribution amplitude
(that can be obtained in the limit of small transverse coordinate)
fulfils the ERBL evolution equation~\cite{Efremov:1979qk,Lepage:1980fj} (a NRQCD
study of the same quantity can be found in \cite{Jia:2008ep,Ma:2006hc,Bell:2008er}). We
are going to show that using this wave function it is possible to
compute NLO corrections to exclusive quarkonium production by analyzing
the divergence structure and checking that it is consistent with B-JIMWLK
evolution \cite{Balitsky:1995ub,Kovchegov:1999yj,Kovchegov:1999ua,Ferreiro:2001qy,Iancu:2000hn,Iancu:2001ad,JalilianMarian:1997dw,JalilianMarian:1997gr,JalilianMarian:1997jx,Kovner:2000pt,Weigert:2000gi}
of the target. A more complete phenomenological analysis will be possible
when the NLO corrections to the photon wave function with massive
quarks, currently under investigation, become available.

The paper is organized as follows. In the next section, we review the
specific formulas of the dipole approach needed for this computation
and fix the notation. In section \ref{sec:wave} we are going to define
the procedure to obtain the non-relativistic wave-function and its
corrections and we are going to compute the relativistic contribution.
In section \ref{sec:Light-cone-distribution-amplitud} we are going
to show that we can use the previous results to obtain the light-cone
distribution amplitude and we are going to check that it satisfies the Efremov-Radyushkin-Brodsky-Lepage (ERBL) evolution equation. Section \ref{sec:Radiative-corrections-to} provides a
strong cross-check of our results by applying them to the computation
of the radiative corrections to S-wave quarkonium decay into leptons.
In section \ref{sec:Exclusive-quarkonium-production} we study the
divergence structure of exclusive quarkonium production in both the
dilute and in the non-dilute limit. Finally, we conclude
in section \ref{sec:Conclusions}.

\section{The dipole approach\label{sec:The-dipole-approach}}

We are going to perform the computations using light-cone coordinates,
for a given vector $p^{\mu}$ they are defined as 
\begin{equation}
p^{+}=\frac{p^{0}+p^{3}}{\sqrt{2}}\qquad p^{-}=\frac{p^{0}-p^{3}}{\sqrt{2}}\,,
\end{equation}
which implies 
\begin{equation}
p^{0}=\frac{p^{+}+p^{-}}{\sqrt{2}}\qquad p^{3}=\frac{p^{+}-p^{-}}{\sqrt{2}}\,.
\end{equation}

As a consequence, the momentum integration measure takes the form
\begin{equation}
d^{4}p=dp^{+}dp^{-}d^{2}p_{\perp}\,,
\end{equation}
and the scalar product of two vectors in these coordinates is
\begin{equation}
p\cdot q=p^{+}q^{-}+p^{-}q^{+}-{\bf p_{\perp}}\cdot {\bf q_{\perp}}\,.
\end{equation}
It is useful to define the light cone vectors 
\begin{equation}
n^{\mu}=\frac{1}{\sqrt{2}}(1,0,0,1)\,,
\end{equation}
and 
\begin{equation}
\bar{n}^{\mu}=\frac{1}{\sqrt{2}}(1,0,0,-1)\,,
\end{equation}
such that
\begin{equation}
n\cdot\bar{n}=1\,,
\end{equation}
\begin{equation}
p^{+}=\bar{n}\cdot p\,,
\end{equation}
\begin{equation}
p^{-}=n\cdot p\,.
\end{equation}
We also find it sometimes useful to separate from four-momenta in the on-shell part, denoted by a hat
\begin{equation}
p^{\mu}=\hat{p}^{\mu}+\bar{p}^{\mu}\,,
\end{equation}
such that 
\begin{equation}
\hat{p}=\left(p^{+},\frac{p_{\perp}^{2}+m^{2}}{2p^{+}},{\bf p_{\perp}}\right)\,.
\end{equation}
In light cone perturbation theory the off-shellness is included in only the ligh-cone energy ($-$ or 
$\bar{n}^{\mu}$-component) of the four vector
\begin{equation}
\bar{p}^{\mu}=\frac{p^{2}-m^{2}}{2p^{+}}\bar{n}^{\mu}\,.
\end{equation}

\begin{figure}
\centerline{\includegraphics[width=0.6\textwidth]{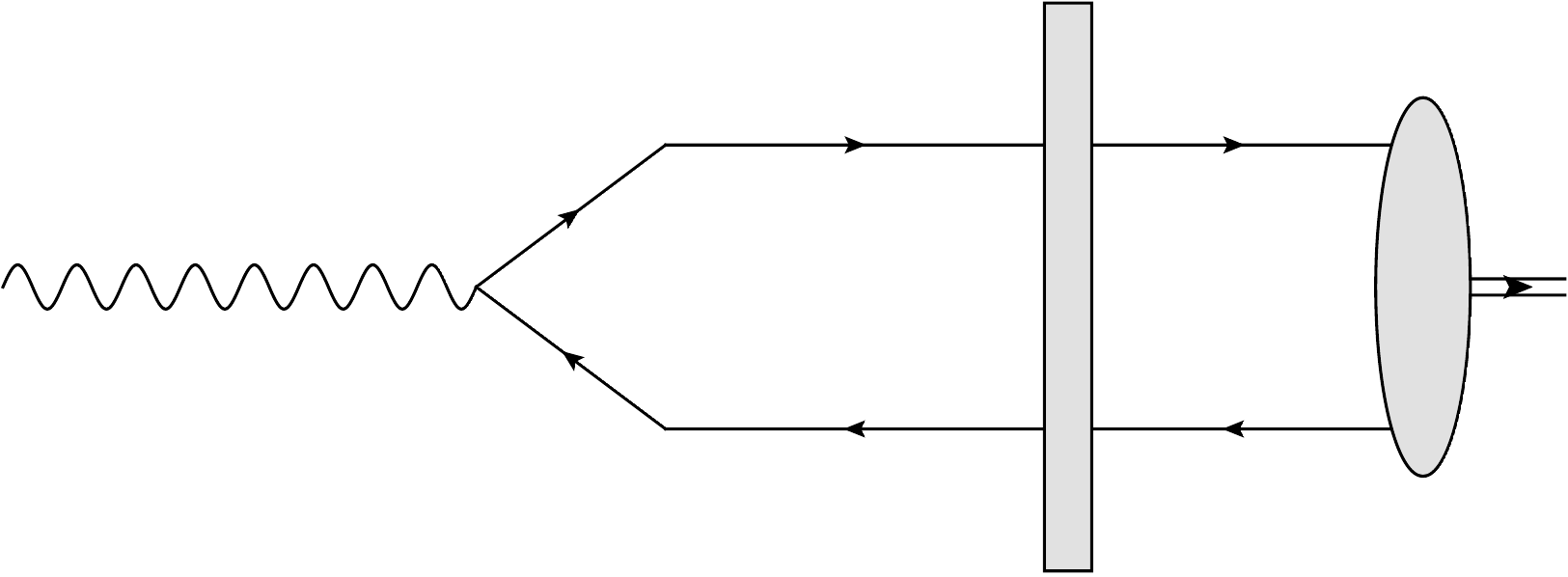}}

\caption{\label{fig:dcross}Representation of the leading order amplitude for
exclusive quarkonium production in the dipole model. A virtual photon
splits into a heavy quark-antiquark pair that later interacts with
the nucleus (represented by a square box). After this interaction,
the pair forms a quarkonium state (represented by a blob). }

\end{figure}

In order to apply the dipole approach to exclusive quarkonium production,
we are going to follow the discussion in Ref.~\cite{Kowalski:2006hc}. We start with eq.
(12) there which corresponds to the leading order contribution, in
the limit of no target recoil $t=0$:
\begin{equation}
\frac{d\sigma_{T,L}^{\gamma^{*}+N\to HQ+N}}{dt}=\frac{1}{16\pi}\left|\int\,d^{2}r_{\perp}\int_{0}^{1}\frac{\,dz}{4\pi}\left(\Psi_{HQ}^{*}\Psi_{\gamma^{*}}\right)_{T,L}\sigma_{q\bar{q}}\right|^{2}\,,
\end{equation}
where the sub-index $T$ or $L$ refers respectively to a transverse or longitudinal polarization of the photon and vector meson~\footnote{The interaction with the target is eikonal, thus the polarization of the meson is the same as that of the photon}. Here $\Psi_{\gamma^{*}}$ and $\Psi_{HQ}$ are respectively
the light-cone wave functions of the photon and quarkonium, depending on the transverse coordinate separation $r_{\perp}$ and $p^ +$ momentum fractions $z,1-z$ of the quark and antiquark in the meson. The properties of the gluon field of the target are encoded in the ``dipole cross section'' $\sigma_{q\bar{q}}$ which is, more properly speaking, twice the imaginary part of the forward elastic scattering amplitude of the quark-antiquark dipole with the target. Importantly for the predictive power of the framework, the same quantity appears in other cross sections, both for DIS and other high energy scattering processes.
A graphical representation can be found in fig. \ref{fig:dcross}.
The previous formula actually assumes that only the subspace of the
Fock space $q\bar{q}$ is relevant for quarkonium light-cone wave
function. A more general formula would involve a sum over all the possible Fock states of the quarkonium:
\begin{equation}
\frac{d\sigma_{T,L}^{\gamma^{*}+N\to HQ+N}}{dt}=\frac{1}{16\pi}\left|\int\,d^{2}r_{\perp}\int_{0}^{1}\frac{\,dz}{4\pi}\left(\Psi_{HQ}^{*}\Psi_{\gamma^{*}}\right)_{T,L}^{q\bar{q}}\sigma_{q\bar{q}}+\int\,d^{2}r_{\perp}\int_{0}^{1}\frac{\,dz}{4\pi}\left(\Psi_{HQ}^{*}\Psi_{\gamma^{*}}\right)_{T,L}^{q\bar{q}g}\sigma_{q\bar{qg}}+\cdots\right|^{2}\,.
\end{equation}

Exclusive quarkonium production is a multi-scale process. In the previous
formula, we identify many different energy scales. Firstly, $Q$ is the virtuality
of the photon and is encoded in its wave-function. The saturation scale of the target $Q_{s}$ is, in the previous formula, hidden inside the $r$-dependence of the amplitudes $\sigma_{q\bar{q}}$ and $\sigma_{q\bar{qg}}$. Physically it represents the typical size of the dipoles selected by the target. Dipoles much smaller than $1/\qs$ interact weakly due to ``color transparency''. For dipoles larger than $1/\qs$ the growth of the amplitude is limited by unitarity, i.e. gluon saturation, and  ceases to compensate for the suppression from the photon and quarkonium wavefunctions. Finally, inside $\Psi_{HQ}$ there are hidden many scales related to quarkonium physics, namely
the scales $m$, $mv$ , $mv^{2}$ and $\Lambda_{QCD}$, which were
discussed before. In this work, we are going to assume that $Q_{s},m\gg mv,\Lambda_{QCD}$
and we are going to explore the physical consequences of this regime.

Details on how to obtain the light-cone wave function of quarkonium
in this approximation will be given in the next section, however,
we can explain the physical implications here. The quarkonium wave function
is dominated by quarks that are moving with a non-relativistic velocity.
This implies that it extends over distances of order $\frac{1}{mv}$
and it is not very sensitive to variations at smaller distance scales.
Another consequence of the non-relativistic nature of quarkonium is
that at leading order $\Psi_{HQ}$ only has support for values of
$z$ around $\frac{1}{2}$. Therefore
\begin{equation}
\frac{d\sigma_{T,L}^{\gamma^{*}+N\to HQ+N}}{dt}\propto\frac{\left|\int\frac{\,d\lambda}{4\pi}\phi^{q\bar{q}}\right|^{2}}{16\pi}\left|\int\,d^{2}r_{\perp}\left(\Psi_{\gamma^{*}}\right)_{T,L}\sigma_{q\bar{q}}\right|_{z=\frac{1}{2}}^{2}\,,\label{eq:lo}
\end{equation}
where $\phi^{q\bar{q}}$ is the leading order light cone wave function
that only takes into account non-relativistic contributions and $\lambda=z-\frac{1}{2}$.
Extending this idea to higher orders, the quarkonium wave function for short distances
${\bf x}_{\perp}\sim\frac{1}{m}$ fulfils
\begin{equation}
\int\,dzf(z)\Psi_{HQ}^{n}(z,{\bf x}_{\perp})=\sum_{m,k}\int\,dzf(z)C_{n\gets m}^{k}(z,{\bf x}_{\perp})\left(\frac{\nabla}{m}\right)^{k}\int\frac{\,d\lambda}{4\pi}\phi^{m}(\lambda,{\bf {\bf 0}})\,.
\end{equation}
In this formula the index $n$ and $m$ represent the particle content; for higher Fock states the single two-particle relative coordinates $z,{\bf x}_{\perp}$ are  replaced by the appropriate variables.  The notation $\Psi_{HQ}$ refers to the full light-cone wave function of quarkonium, while the $\phi$'s are
the wavefunctions restricted to the case in which all particles are non-relativistic
(therefore $\lambda\ll1$). Here $f(z)$ is just a test function written
to represent the fact that the equality is only valid when integrating
for $z$. At this stage, we are ignoring spin degrees of freedom in the notation,
but we will be more explicit about this in the following sections.
Note that each power of $\left(\frac{\nabla}{M}\right)$ acting on
$\phi$ gives a suppression of order $v$ (in this equation $\nabla$ is a three-vector, such that $\nabla=\nabla_\perp+i2m\lambda$). Due to the non-relativistic
nature of quarkonium the coefficients start with just a heavy quark-antiquark-state which is, to leading order, nonrelativistic:
\begin{equation}
C_{q\bar{q}\gets q\bar{q}}^{0}=4\pi\delta(z-\frac{1}{2})+\mathcal{O}(\alpha_{s})\,,
\end{equation}
with transitions to higher Fock states entering at higher orders in perturbation theory. 
\begin{equation}
C_{n\gets m}^{0}|_{n\neq m}=0+\mathcal{O}(\alpha_{s})\,.
\end{equation}

\begin{figure}
a)\includegraphics[scale=0.5]{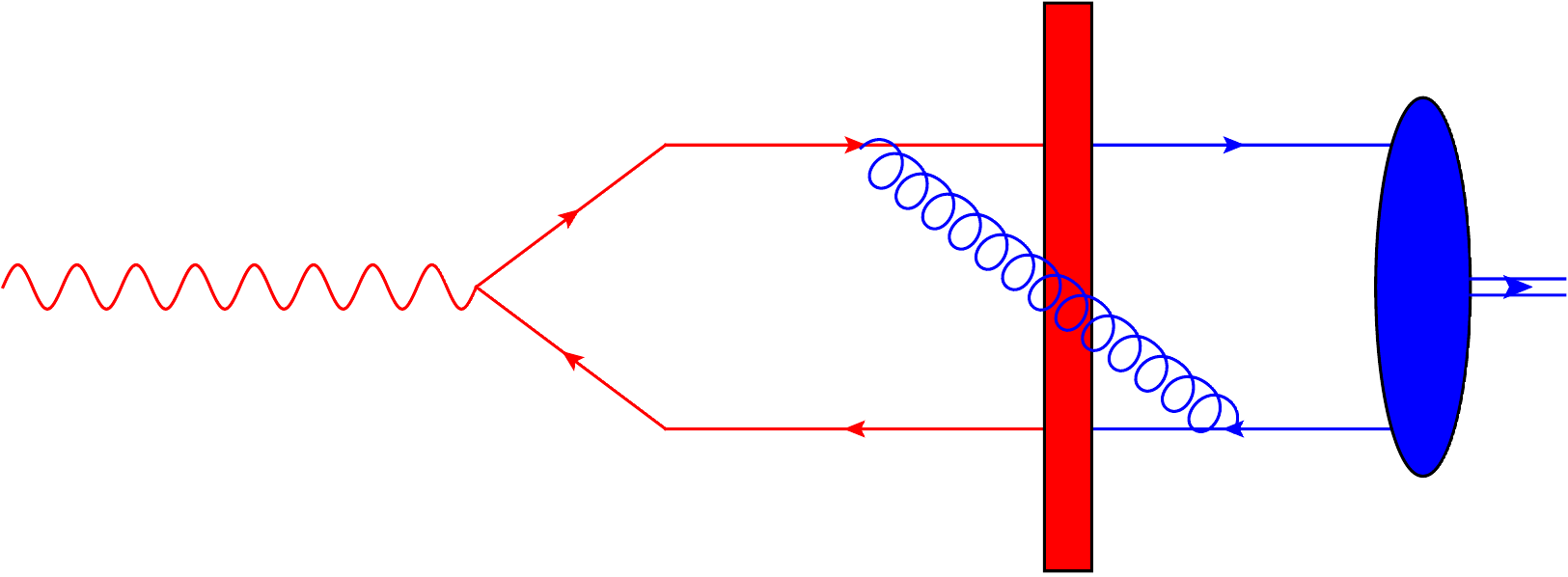} b)\includegraphics[scale=0.5]{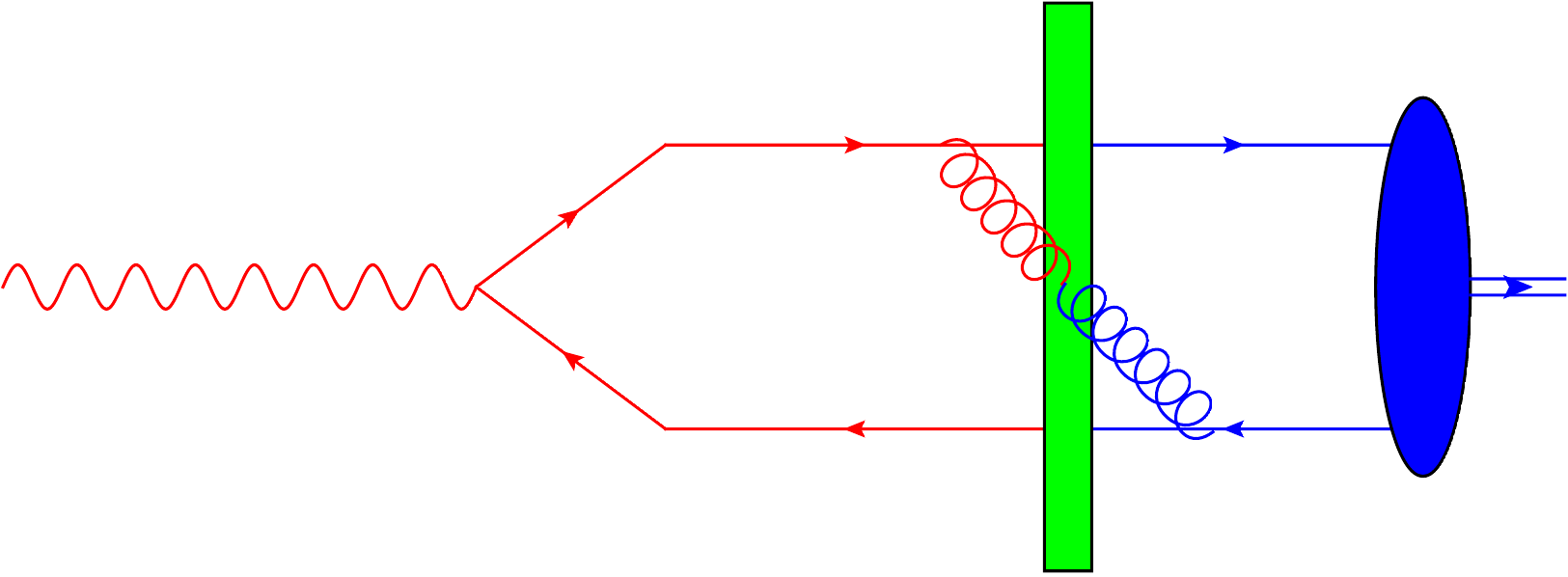}

\caption{\label{fig:nrvsdi}Comparison between how the degrees of freedom are
separated in NRQCD and in the dipole model for the particular case
in which the heavy quark pair emits a soft gluon before crossing the
target. a) In NRQCD degrees of freedom are separated according to
their virtuality. In this picture, red particles have a virtuality
larger than $m^{2}v^{2}$ (we are assuming, like in the rest of the
text, that $Q_{s}\gg mv$) and their influence would be encoded in
the hard coefficient. The rest of the particles are represented in
blue and will influence the NRQCD matrix elements. b) In the dipole
model degrees of freedom are separated according to time. Particles
before crossing the target belong to the wave function of the photon
and we represent them here in red while particles after the target
are encoded in the wave function of quarkonium and are coloured blue.
The method that we are implementing in this manuscript combines
the two schemes by separating relativistic from non-relativistic degrees
of freedom in the wave function of quarkonium.}

\end{figure}

In this paper, we are going to argue that in order to compute exclusive
quarkonium production at next-to-leading order (NLO) only $C_{q\bar{q}\gets q\bar{q}}^{0}$
and $C_{q\bar{q}g\gets q\bar{q}}^{0}$ at order $\alpha_{s}$ are needed
and we are going to compute them. The fact that the contributing Fock states at NLO are the 
$q\bar{q}$ and $q\bar{q}g$ ones is similar to other NLO calculations in the dipole picture (e.g.\cite{Beuf:2016wdz,Beuf:2017bpd,Hanninen:2017ddy,Boussarie:2016ogo,Boussarie:2016bkq}). The additional feature in the case of quarkonium is the way how both are related to a common nonrelativistic bound state wave function. 
Taking into  account these two Fock states we can write the cross-section as
\begin{equation}
\frac{d\sigma_{T,L}^{\gamma^{*}+N\to HQ+N}}{dt}=\frac{1}{16\pi}\left|\sum_{n,m,k}\left(\left(\frac{\nabla}{M}\right)^{k}\int\frac{\,d\lambda}{4\pi}\phi^{m}(\lambda,{\bf {\bf 0}})\right)\int\,d^{2}r_{\perp}\int_{0}^{1}\frac{\,dz}{4\pi}\left(\left(C_{n\gets m}^{k}(z,{\bf r}_{\perp})\right)^{*}\Psi_{\gamma^{*}}\right)_{T,L}^{n}\sigma_{n}\right|^{2}\, . \label{eq:Cexpand}
\end{equation}
These terms can be reorganized in such a way that we have an expansion
that resembles a typical production process in NRQCD
\begin{multline}
\frac{d\sigma_{T,L}^{\gamma^{*}+N\to HQ+N}}{dt}
\\
=\frac{1}{16\pi}\sum_{n,m,k,n',m',k'}\left(\int\,d^{2}r_{\perp}\int_{0}^{1}\frac{\,dz}{4\pi}\left(\left(C_{n\gets m}^{k}(z,{\bf r}_{\perp})\right)^{*}\Psi_{\gamma^{*}}\right)_{T,L}^{n}\sigma_{n}\int\,d^{2}r'_{\perp}\int_{0}^{1}\frac{\,dz'}{4\pi}\left(C_{n'\gets m'}^{k'}(z',{\bf r}'_{\perp})\Psi_{\gamma^{*}}^{*}\right)_{T,L}^{n'}\sigma*_{n'}\right)
\\
\times 
\left(\left(\frac{\nabla}{m}\right)^{k}\int\frac{\,d\lambda}{4\pi}\phi^{m}(\lambda,{\bf {\bf 0}})\left(\frac{\nabla}{m}\right)^{k'}\int\frac{\,d\lambda'}{4\pi}\phi^{m'}(\lambda',{\bf {\bf 0}})\right)\,,\label{eq:nrqcdlike}
\end{multline}
In this equation, the third line only encodes what in NRQCD would
be called soft physics (scales smaller than $m$), however, the second
line receives contributions from both hard and soft physics. For example,
there will be a soft physics contribution in the photon wave-function
whenever a soft gluon is emitted before crossing the target. Therefore
the analogy with NRQCD is not complete. This is illustrated with an
example in fig. \ref{fig:nrvsdi}. Note also that the dipole picture
is formulated in light-cone gauge and therefore we cannot, in principle,
use gauge invariance arguments that are common in NRQCD and that will
impose relations between the different terms in the expansion. For
example, the fact that derivatives acting on the low energy matrix
elements have to be combined with gauge fields to form covariant derivatives
connects states with a different number of particles in such a way
that we would expect that $C_{q\bar{q}\gets q\bar{q}}^{1}$ is constrained
by $C_{q\bar{q}\gets q\bar{q}g}^{0}$, however since we are using a
setting that is not gauge invariant we are not going to use this kind
of relations (for a more complete discussion see \cite{Hoodbhoy:1996zg}, and also the discussion about genuine and kinematical twist in \cite{Altinoluk:2019fui}).

Let us now discuss the power counting of this computation and which
terms do we need to consider if we want to achieve a NLO result. In
this problem radiative corrections in the hard part will be suppressed
by powers of $\alpha_{s}(\mu)|_{\mu\geq m}$, each additional $\frac{\nabla}{m}$
acting on $\phi$ would amount for a suppression of order $v\sim\alpha_{s}\left(mv\right)$.
From now on we will ignore, for purposes of order estimation, the
scale at which $\alpha_{s}$ is evaluated. With this in mind, we should
in principle take into account the first radiative correction in the
hard part and the first velocity correction in the soft part. However,
the first velocity correction only enters at NNLO. This is because due to rotational
invariance the contribution from $k=1$ and $k'=0$ (or vice-versa)
in \eq\nr{eq:nrqcdlike} is zero. At LO the only contributing Fock state is
$n=m=q\bar{q}$ (and
correspondingly in the complex conjugate amplitude denoted by primes). 
At NLO $n$ and $m$ can also have the
value $q\bar{qg}$. However, we can also check that the case $m$
(or $m'$) equal to $q\bar{q}g$ is further suppressed. The reason is
that the soft gluon has to be emitted either from the photon wave
function (before the target) or by the relativistic part of the quarkonium
wave function. In both cases, it would correspond to the emission
of a soft gluon by a particle with a virtuality of order $m^{2}$
or higher, and this process is always suppressed by powers of $v$.
In summary, to obtain a NLO result we only need to consider the following
formula

\begin{equation}
\frac{d\sigma_{T,L}^{\gamma^{*}+N\to HQ+N}}{dt}=\frac{\left|\int\frac{\,d\lambda}{4\pi}\phi^{q\bar{q}}\right|^{2}}{16\pi}\left|\int\,d^{2}r_{\perp}\int_{0}^{1}\frac{\,dz}{4\pi}\left(\left(C_{q\bar{q}\gets q\bar{q}}^{0}\right)^{*}\Psi_{\gamma^{*}}^{q\bar{q}}\sigma_{q\bar{q}}+\left(C_{q\bar{q}g\gets q\bar{q}}^{0}\right)^{*}\Psi_{\gamma^{*}}^{q\bar{q}g}\sigma_{q\bar{q}g}\right)\right|^{2}\,.
\label{eq:pcounting}
\end{equation}

Here we have neglected gradient corrections  (higher values of $k,k'$ in \eq\nr{eq:Cexpand}), since in the strict weak coupling power counting, valid  for $mv\gg\Lambda_{QCD}$ the quark velocity is $v\sim\alpha_{s}(mv)$, and thus $v^{2} \ll \alpha_{s}(m)$. It is worth keeping in mind that for practical applications in the
case of charmonium it might happen that numerically $v^{2}$ can be comparable to $\alpha_{s}(m)$, making the gradient corrections important. Nevertheless, it is always true that the NLO radiative corrections are bigger or of the order of the  first corrections due to the velocity expansion and, therefore, a requisite of any improvement over the LO result.

The power counting discussed here is, to our knowledge, not clearly expressed in the literature, where typically only the LO limit is considered. Thus \eq\nr{eq:pcounting} is the first main result of this paper. To make this statement quantitative, we will next proceed to calculate at NLO the coefficient functions $C_{q\bar{q}\gets q\bar{q}}^{0}$
and $C_{q\bar{q}g\gets q\bar{q}}^{0}$. We will then show that the cross section
formula \nr{eq:pcounting} gives finite results for any $\sigma_{q\bar{q}}$ and $\sigma_{q\bar{q}g}$
that fulfil B-JIMWLK evolution.
However, it is also interesting to discuss the relation with collinear factorization
\cite{Lepage:1980fj,Chernyak:1983ej} that can be used when $Q$ is
larger than any other scale in the problem. In that formalism the
cross-section can be understood as the convolution of a hard function
with the light-cone distribution amplitude (LCDA), which in the light-cone
gauge corresponds to the light-cone wave function in the limit ${\bf r}_{\perp}\to{\bf 0}$.
The LCDA must fulfil the ERBL evolution \cite{Efremov:1979qk,Lepage:1980fj}.
In \cite{Jia:2008ep,Ma:2006hc,Bell:2008er} the LCDA and ERBL evolution were studied within
the formalism of NRQCD. We are going to check that LCDA that we can
obtain from our light-cone wave function fulfils ERBL evolution. 

\section{The wave function of heavy quarkonium}
\label{sec:wave}

\begin{figure}[tbh!]
\centerline{\includegraphics[width=0.5\textwidth]{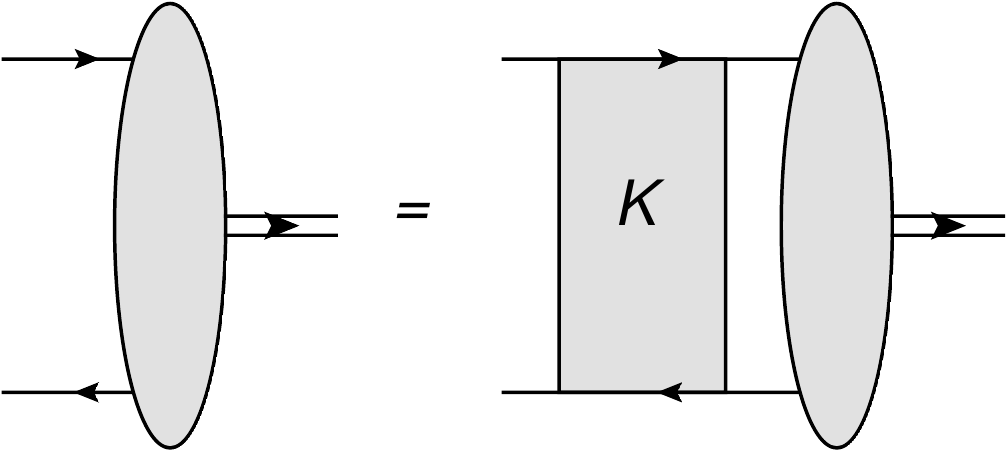}}
\caption{\label{fig:bethenr}Bethe-Salpeter equation that gives the non-relativistic
wave function. In this picture all particles are non-relativistic,
therefore, the kernel $K$ is just the Green function of the scattering
of two non-relativistic heavy quarks.}
\end{figure}

The two body part of the wave function of quarkonium
can be determined using the Bethe-Salpeter equation. In the non-relativistic
limit, the relation between the light-cone wave function and the wave
function that can be obtained in a potential model was studied in
\cite{Bodwin:2006dm}. Here we use this result as a starting point.
Following \cite{Kowalski:2006hc}, we use the convection that the
light cone wave function is normalized such that 
\begin{equation}
\sum_{h\bar{h}}\int\,d^{2}r_{\perp}\int_{0}^{1}\frac{\,dz}{4\pi}|\phi_{h\bar{h}}(z,r_{\perp})|^{2}=1\,,\label{eq:normlc}
\end{equation}
where in this formula we have written explicitly the sum over polarizations.
We are going now to analyze in more detail the LO light-cone wave
function of a vector meson. Following \cite{Bodwin:2006dm} we get
in the spinor matrix space (instead of helicity space) wave function
\begin{equation}
\phi_{ab}^{\mu}(z,r_{\perp})=\lambda\tilde{\phi}(z,r_{\perp})\left(\frac{1+\slashed{v}}{2}\gamma^{\mu}\frac{1-\slashed{v}}{2}\right)_{ab}\,,
\end{equation}
where $\tilde{\phi}$ is a function normalized as it is usually done
in the context of potential models
\begin{equation}
\int\,d^{3}x|\tilde{\phi}(x)|^{2}=1\,,\label{eq:normpm}
\end{equation}
 and $\lambda$ is a constant that we are going to fix such that eq.
(\ref{eq:normlc}) follows from eq. (\ref{eq:normpm}). We can transform
the last expression to light cone helicity space  using the prescription
in \cite{Lepage:1980fj}, obtaining 
\begin{equation}
\phi_{h\bar{h}}^{\mu}(z,r_{\perp})=-\frac{\lambda\tilde{\phi}(z,r_{\perp})}{8m^{2}}\bar{u}(mv,h)\gamma^{\mu}v(mv,\bar{h})\,.
\end{equation}
The normalization condition then imposes that $\lambda=\sqrt{32m^{3}}$
so we finally obtain for the longitudinal component
\begin{equation}
\phi_{h\bar{h}}^{3}(z,r_{\perp})=-\sqrt{2m}\tilde{\phi}(z,r_{\perp})\delta_{h,-\bar{h}}\,,
\end{equation}
while for the transverse component, we get 
\begin{equation}
\phi_{h\bar{h}}^{i}(z,r_{\perp})=\sqrt{2m}\tilde{\phi}(z,r_{\perp})\delta_{h\bar{h}}(\delta_{i1}(-1)^{\frac{1-h}{2}}-i\delta_{i2})\,.
\end{equation}
The previous functions fulfil the Bethe-Salpeter equation represented in fig.
\ref{fig:bethenr} which can equivalently be written as a Schrödinger
equation~\cite{Bodwin:2006dm}.

\begin{figure}[tbh!]
\centerline{\includegraphics[width=0.3\textwidth]{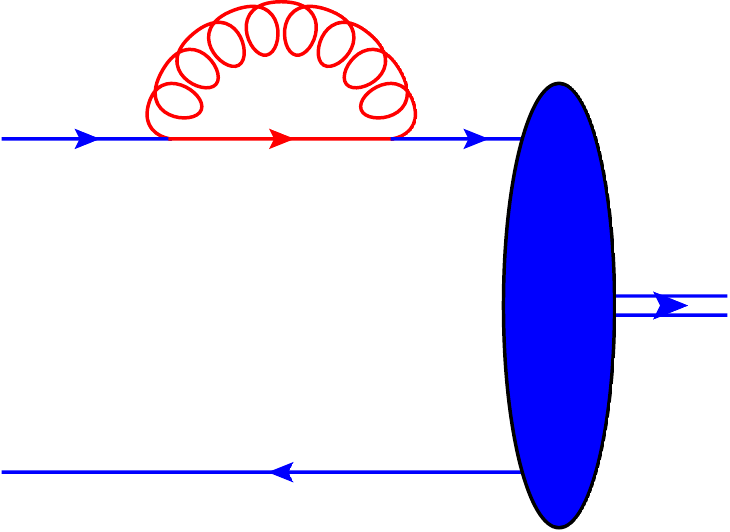}}
\caption{\label{fig:bethese} Example of a contribution that can be encoded
in the wave function renormalization of a non-relativistic quark.
We use the same colour code as in part a) of fig. \ref{fig:nrvsdi}.}
\end{figure}

\begin{figure}[tbh!]
\centerline{\includegraphics[width=0.3\textwidth]{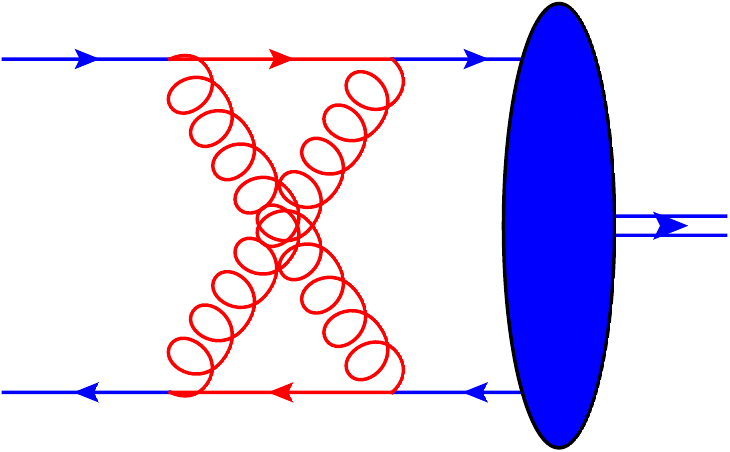}}
\caption{\label{fig:kernelr} Example of a contribution that can be encoded
as a correction to the kernel.}
\end{figure}

Now we want to go beyond the non-relativistic approximations and consider
the influence of relativistic quarks and gluons with an energy of
the order of the heavy quark mass. We can differentiate three cases:
\begin{enumerate}
\item Components of the wave function that correspond
to the probability amplitude of having a relativistic particle inside
the quarkonium state. 
\item Contributions (from relativistic particles in loops) resulting in a wave function renormalization of a non-relativistic particle of which the probability amplitude
to be in a quarkonium state is considered. An example of  such a contribution is shown in fig. \ref{fig:bethese}.
\item Contributions that can be encoded as a correction to the kernel $K$
in fig. \ref{fig:bethenr}. An example is shown in fig. \ref{fig:kernelr}.
\end{enumerate}
We will not discuss contributions of type 3. Their effect can be encoded
in a redefinition of the potential (or equivalently, the kernel in
the Bethe-Salpeter equation). Hence, in the case of production and
decay processes this information is hidden in the value of the non-relativistic
wave function at the origin. Contributions of type 2 are important
but they just renormalize multiplicatively the non-relativistic wave
function by a constant equal to the wave function renormalization\footnote{Exterior heavy quark lines in the Bethe-Salpeter equation are multiplied
by $\sqrt{Z}$. Since there are two of them the overall factor is
$Z$. }. At leading order they correspond to the diagram in \ref{fig:bethese} which  we compute it in appendix \ref{sec:The-wave-function-renormalizatio},
giving
\begin{align}
Z= & 1+\delta Z+\mathcal{O}(\alpha_{s}^{2})\,, \\ \nonumber
\delta Z= & -\frac{\alpha_{s}C_{F}}{2\pi}
\bigg[\frac{1}{D-4}(4\log x_{0}+3)+2\log x_{0}\left(\log\left(\frac{m^{2}}{4\pi\mu^{2}}\right)+1+\log x_{0}\right)
\\ & \quad \quad
+(4\log x_{0}+3)\frac{\gamma_{E}}{2}+\frac{3}{2}\log\left(\frac{m^{2}}{4\pi\mu^{2}}\right)-2\bigg]\,.
\label{eq:Z}
\end{align}

In this last expression, we have regulated transverse momentum in
dimensional regularization such that the dimensions of the transverse
space is $D-2$. The $+$-component of the momentum has been regulated
using an infrared cut-off $x_{0}$ in the momentum fraction of the
gluon (with respect to the initial quark). This is a similar regularization scheme as used in the NLO LCPT calulations in~\cite{Beuf:2016wdz,Hanninen:2017ddy,Beuf:2017bpd}.  The wave function renormalization
that we obtained is similar to the one presented in \cite{Mustaki:1990im}
but not exactly equal. However, our results agree on the double-logarithm
contribution.

Contributions of type 1 can be computed by considering diagrams in
which all relativistic particles annihilate to form non-relativistic
ones. They will be proportional to  the non-relativistic wave function
at the origin. The reason for this simple structure is that in diagrams
in which only relativistic particles are involved the only energy
scale that appears is $m$. Since no momenta are parametrically of order $mv$ or $mv^2$, there are no diagrams enhanced by inverse powers of the velocity $v$, which would require a resummation. Furthermore, since all relativistic processes are short distance processes from
the point of view of non-relativistic wave function, it is a good approximation
(up to additional powers of $v$) to consider that the non-relativistic
particles are created at the same point. In summary, contributions
of types 1 and 2 will be encoded in the functions $C_{n\gets m}^{k}$
that were introduced in the previous sections. The difference is that
while contributions of type 2 give rise to corrections that are just
constants, contributions of type 1 will have a non-trivial dependence
on $r_{\perp}$ and $z$. On the other hand, contributions of type
3 will be encoded in the non-relativistic wave function.

\begin{figure}[tb!]
\centerline{\includegraphics[width=0.5\textwidth]{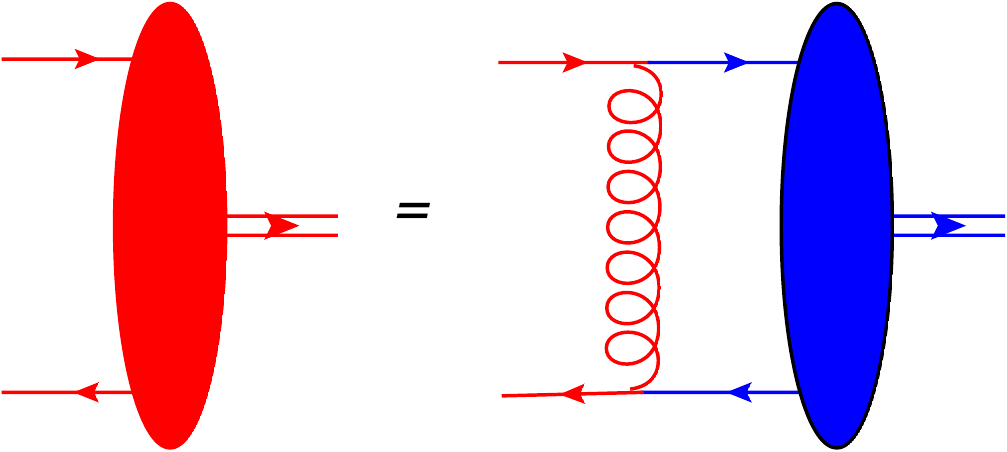}}

\caption{\label{fig:kernelr-1} LO contribution to the component of the wave
function with two relativistic heavy quarks.}

\end{figure}

For the computation at hand, we are interested in the components of
the light-cone wave function in which we have two relativistic quarks
or a relativistic quark, a non-relativistic one and a hard gluon.
Let us focus first of the former case. Its LO contribution is represented
by the diagram in fig. \ref{fig:kernelr-1}. Details of the computation
will be given in appendix \ref{sec:Relativistic-correction-to}, the
result from this contribution gives
\begin{align}
\left.\left\{ \Psi_{HQ}^{q\bar{q}}\right\} _{\lambda_{1}\lambda_{2}}^{i}(z,p_{\perp})\right|_{Fig.\ref{fig:kernelr-1}}= & \frac{4g^{2}C_{F}\sqrt{z(1-z)}}{p_{\perp}^{2}+4m^{2}(z-\frac{1}{2})^{2}}\left(\frac{d_{\mu\nu}(\hat{p})(\theta(z-\frac{1}{2})(1-z)+\theta(\frac{1}{2}-z)z)}{p_{\perp}^{2}+4m^{2}(z-\frac{1}{2})^{2}}+\frac{\bar{n}_{\mu}\bar{n}_{\nu}}{4m^{2}(z-\frac{1}{2})^{2}}\right)\times\nonumber \\
 & \times\sum_{\lambda_{1}',\lambda_{2}'}[\bar{u}(\hat{p}_{Q},\lambda_{1})\gamma^{\mu}u(mv,\lambda_{1}')]\int\frac{\,d\lambda}{4\pi}\phi_{q\bar{q}}^{i}(\lambda,{\bf {\bf 0}})[\bar{v}(mv,\lambda_{2}')\gamma^{\nu}v(\hat{p}_{\bar{Q}},\lambda_{2})]\,,\label{eq:psirel}
\end{align}
where $p$ is the momentum exchanged by the gluon and the light cone gauge gluon polarization sum is
\begin{equation}
d_{\mu\nu}(q)=g_{\mu\nu}-\frac{\bar{n}_{\mu}q_{\nu}+q_{\mu}\bar{n}_{\nu}}{\bar{n}\cdot q}=-\sum_{\lambda_{G}}\epsilon^{\mu}(q,\lambda_{G})\epsilon^{\nu}(q,\lambda_{G})\,.
\end{equation}

Note also that the function $\phi_{q\bar{q}}^{i}(\lambda,{\bf {\bf 0}})$
has to be understood as $\phi_{q\bar{q}}^{i}(\lambda,{\bf {\bf {\bf r}_{\perp}=0}})$.
For future manipulations, it might be useful to write this result
in coordinate space. To do this we make use of the formulas in Appendix
\ref{sec:Dirac-algebra-manipulations} to rewrite
\begin{multline}
\bar{u}(\hat{p}_{Q},\lambda_{1})\slashed{\epsilon}u(mv,\lambda_{1}')=\frac{p_{\perp}^{i}\epsilon_{\perp}^{j}}{2\sqrt{2}mz\left(z-\frac{1}{2}\right)}\bar{u}(\hat{p}_{Q},\lambda_{1})\slashed{\bar{n}}\left[\left(z+\frac{1}{2}\right)\delta^{ij}+\frac{z-\frac{1}{2}}{2}[\gamma_{\perp}^{i},\gamma_{\perp}^{j}]\right]u(mv,\lambda_{1}')
\\
+\frac{z-\frac{1}{2}}{\sqrt{2}z}\bar{u}(\hat{p}_{Q},\lambda_{1})\slashed{\epsilon}_{\perp}\slashed{\bar{n}}u(mv,\lambda_{1}')\,,
\label{eq:ubarepslashu}
\end{multline}
and
\begin{multline}
\bar{v}(mv,\lambda_{2}')\slashed{\epsilon}v(\hat{p}_{\bar{Q}},\lambda_{2})  =\frac{p_{\perp}^{i}\epsilon_{\perp}^{j}}{2\sqrt{2}m(1-z)\left(z-\frac{1}{2}\right)}\bar{v}(mv,\lambda_{2}')\slashed{\bar{n}}\left[\left(\frac{3}{2}-z\right)\delta^{ij}+\frac{z-\frac{1}{2}}{2}[\gamma_{\perp}^{i},\gamma_{\perp}^{j}]\right]v(\hat{p}_{\bar{Q}},\lambda_{2})
\\
  -\frac{z-\frac{1}{2}}{\sqrt{2}(1-z)}\bar{v}(mv,\lambda_{2}')\slashed{\epsilon}_{\perp}\slashed{\bar{n}}v(\hat{p}_{\bar{Q}},\lambda_{2})\,.
\label{eq:vbarepslashv}
\end{multline}

Using the formulas of Appendix \ref{sec:Useful-Fourier-transforms} and the expressions for the spinor matrix elements in Appendix~\ref{sec:fig6_intermediate}
it is possible to find the result in coordinate space 
\begin{align}
\left.\left\{ \Psi_{HQ}^{q\bar{q}}\right\} _{\lambda_{1}\lambda_{2}}^{i}(z,r_{\perp})\right|_{Fig.\ref{fig:kernelr-1}} & =\frac{2g^{2}C_{F}z(1-z)}{\pi \left(z-\frac{1}{2}\right)^{2}}\left[K_{0}(\tau)-\frac{\left(z+\frac{1}{2}\right)\left(\frac{3}{2}-z\right)}{2z(1-z)}(\theta(z-\frac{1}{2})(1-z)+\theta(\frac{1}{2}-z)z)\left(K_{0}(\tau)-\frac{\tau}{2}K_{1}(\tau)\right)\right]\times\nonumber \\
 & \times\int\frac{\,d\lambda}{4\pi}\left\{\phi_{q\bar{q}}^{i}(\lambda,{\bf {\bf 0}})\right\}_{\lambda_1,\lambda_2}\nonumber \\
 & -\frac{g^{2}C_{F}}{\pi}(\theta(z-\frac{1}{2})(1-z)+\theta(\frac{1}{2}-z)z)K_{0}(\tau)(-1)^{\frac{2-\lambda_1-\lambda_2}{2}}\left\{\int\frac{\,d\lambda}{4\pi}\phi_{q\bar{q}}^{i}(\lambda,{\bf {\bf 0}})\right\}_{\lambda_1,\lambda_2}\nonumber \\
 & +\frac{g^{2}C_{F}\left(z-\frac{1}{2}\right)mr_\perp}{\pi}(\theta(z-\frac{1}{2})(1-z)+\theta(\frac{1}{2}-z)z)K_{1}(\tau)\times(-1)^{\frac{2-\lambda_1-\lambda_2}{2}}\left\{\int\frac{\,d\lambda}{4\pi}\phi_{q\bar{q}}^{i}(\lambda,{\bf {\bf 0}})\right\}_{\lambda_1,\lambda_2}\nonumber \\
 & -\frac{ig^{2}C_{F}}{2\left(z-\frac{1}{2}\right)\pi }(\theta(z-\frac{1}{2})(1-z)+\theta(\frac{1}{2}-z)z)\tau K_{0}(\tau)\left(\left(z+\frac{1}{2}\right)-(-1)^{\frac{2-\lambda_1-\lambda_2}{2}}\left(z-\frac{1}{2}\right)\right)\times\nonumber \\
 & \times\left\{\int\frac{\,d\lambda}{4\pi}\phi_{q\bar{q}}^{i}(\lambda,{\bf {\bf 0}})\right\}_{\lambda_1,-\lambda_2}\frac{r_{\perp}^{j}}{r_\perp}((-1)^{\frac{1-\lambda_2}{2}}\delta^{j1}-i\delta^{j2})\nonumber \\
 & +\frac{ig^{2}C_{F}}{2\left(z-\frac{1}{2}\right)\pi }(\theta(z-\frac{1}{2})(1-z)+\theta(\frac{1}{2}-z)z)\tau K_{0}(\tau)\left(\left(\frac{3}{2}-z\right)+(-1)^{\frac{2-\lambda_1-\lambda_2}{2}}\left(z-\frac{1}{2}\right)\right)\times\nonumber \\
 & \left\{\int\frac{\,d\lambda}{4\pi}\phi_{q\bar{q}}^{i}(\lambda,{\bf {\bf 0}})\right\}_{-\lambda_1,\lambda_2}\frac{r_\perp^j}{r_\perp}((-1)^{\frac{1-\lambda_1}{2}}\delta^{j1}-i\delta^{j2})\nonumber \\
 & +\frac{g^{2}C_{F}\left(z-\frac{1}{2}\right)mr_{\perp}}{\pi}(\theta(z-\frac{1}{2})(1-z)+\theta(\frac{1}{2}-z)z)K_{1}(\tau)\times\nonumber\\
 &\times\left((-1)^{\frac{2-\lambda_1-\lambda_2}{2}}-1\right)\left\{\int\frac{\,d\lambda}{4\pi}\phi_{q\bar{q}}^{i}(\lambda,{\bf {\bf 0}})\right\}_{-\lambda_1,-\lambda_2}\, ,
 \label{eq:longwf}
\end{align}
where $\tau=2m\left(z-\frac{1}{2}\right)r_{\perp}$.

The general expression~ \nr{eq:longwf} can be further simplified if we distinguish the cases
of longitudinal and transverse polarization. For the longitudinal polarization
state we obtain 
\begin{multline}
\label{eq:Lwf}
\left.\left\{ \Psi_{HQ}^{q\bar{q}}\right\} _{\lambda_{1}\lambda_{2}}^{3}(z,r_{\perp})\right|_{Fig.\ref{fig:kernelr-1}}=-\frac{2\sqrt{2m}g^2C_Fz(1-z)\delta_{\lambda_1,-\lambda_2}}{\pi\left(z-\frac{1}{2}\right)}\bigg\{K_{0}(\tau)
\\
+\left(\theta(z-\frac{1}{2})(1-z)+\theta(\frac{1}{2}-z)z\right)
\bigg[\frac{\left(z-\frac{1}{2}\right)^{2}-\frac{1}{2}}{z(1-z)}\left(K_{0}(\tau)-\frac{\tau}{2}K_{1}(\tau)\right)
-\frac{\left(z-\frac{1}{2}\right)^{2}}{2z(1-z)}\tau K_{1}(\tau)\bigg]\bigg\}
\int\frac{\,d\lambda}{4\pi}\tilde{\phi}_{q\bar{q}}(\lambda,{\bf {\bf 0}})\,.
 \end{multline}
The one-loop wave function for the transverse polarization state is
\begin{multline}
\label{eq:Twf}
\left.\left\{ \Psi_{HQ}^{q\bar{q}}\right\} _{\lambda_{1}\lambda_{2}}^{i}(z,r_{\perp})\right|_{Fig.\ref{fig:kernelr-1}}
\\
 =\frac{2g^{2}C_{F}z(1-z)}{\pi\left(z-\frac{1}{2}\right)^{2}}\left[K_{0}(\tau)-\frac{(\theta(z-\frac{1}{2})(1-z)+\theta(\frac{1}{2}-z)z)}{2z(1-z)}\left(K_{0}(\tau)-\frac{\tau}{2}K_{1}(\tau)\right)\right]
 \\ \shoveright{\times
 \sqrt{2m}\delta_{\lambda_1,\lambda_2}(\delta_{i1}(-1)^{\frac{1-\lambda_1}{2}}-i\delta_{i2})\int\frac{\,d\lambda}{4\pi}\tilde{\phi}_{q\bar{q}}(\lambda,{\bf {\bf 0}})
}
 \\
 +\frac{i2g^2C_F\tau K_0(\tau)}{\pi}\left(\theta(z-\frac{1}{2})(1-z)+\theta(\frac{1}{2}-z)z\right)
 \\ \times
 \delta_{\lambda_1,-\lambda_2}\frac{r^j}{r}\left(\delta^{ij}+\frac{i(-1)^{\frac{1-\lambda_1}{2}}\epsilon^{ij}}{2\left(z-\frac{1}{2}\right)}\right)\sqrt{2m}\int\frac{\,d\lambda}{4\pi}\tilde{\phi}_{q\bar{q}}(\lambda,{\bf {\bf 0}})
\end{multline}

Combining the results of eq. \nr{eq:Z} and \nr{eq:psirel},
we obtain that
\begin{multline}
C_{q\bar{q}\gets q\bar{q}}^{0}(z,{\bf p}_{\perp};\lambda_{1},\lambda_{2};\lambda_{1}',\lambda_{2}')  =4\pi\delta(z-\frac{1}{2})(1+\delta Z)\delta_{\lambda_{1}\lambda_{1}'}\delta_{\lambda_{2}\lambda_{2}'}
\\
  +\frac{4g^{2}C_{F}\sqrt{z(1-z)}}{p_{\perp}^{2}+4m^{2}(z-\frac{1}{2})^{2}}\left(\frac{d_{\mu\nu}(\hat{p})(\theta(z-\frac{1}{2})(1-z)+\theta(\frac{1}{2}-z)z)}{p_{\perp}^{2}+4m^{2}(z-\frac{1}{2})^{2}}+\frac{\bar{n}_{\mu}\bar{n}_{\nu}}{4m^{2}(z-\frac{1}{2})^{2}}\right)
   \\
  \times\sum_{\lambda_{1}',\lambda_{2}'}[\bar{u}(\hat{p}_{Q},\lambda_{1})\gamma^{\mu}u(mv,\lambda_{1}')][\bar{v}(mv,\lambda_{2}')\gamma^{\nu}v(\hat{p}_{\bar{Q}},\lambda_{2})]+\mathcal{O}(\alpha_{s}^{2})\,.\label{eq:qbarqmom}
\end{multline}
This is the first major result of this section. An explicit expression for the structures in \nr{eq:qbarqmom} in coordinate space and for the meson polarization states can be easily obtained using \nr{eq:Lwf} and \nr{eq:Twf}. The result \nr{eq:qbarqmom} represents one of the two NLO corrections to the heavy quarkonium wavefunction, needed for the NLO calculation of exclusive quarkonium production.

In the non-relativistic limit the polarization of the meson is reflected only in the spin state of the quarks, and the spatial part of the wavefunction becomes rotationally symmetric. We can utilize this simplification and integrate the wavefunction over the polar angle, which results in simpler expressions. The general azimuthally symmetric part of the wavefunction is
\begin{align}
\int\,d\theta_{r} & \left.\left\{ \Psi_{HQ}^{q\bar{q}}\right\} _{\lambda_{1}\lambda_{2}}^{i}(z,r_{\perp})\right|_{Fig.\ref{fig:kernelr-1}}
=
\\ \nonumber &
\frac{g^{2}C_{F}\sqrt{z(1-z)}}{m^{2}\left(z-\frac{1}{2}\right)^{2}}\left[K_{0}(\tau)-\frac{\left(z+\frac{1}{2}\right)\left(\frac{3}{2}-z\right)}{2z(1-z)}(\theta(z-\frac{1}{2})(1-z)+\theta(\frac{1}{2}-z)z)\left(K_{0}(\tau)-\frac{\tau}{2}K_{1}(\tau)\right)\right]
\\ \nonumber &
\quad  \quad\times\sum_{\lambda_{1}',\lambda_{2}'}[\bar{u}(\hat{p}_{Q},\lambda_{1})\slashed{\bar{n}}u(mv,\lambda_{1}')]\int\frac{\,d\lambda}{4\pi}\phi_{q\bar{q}}^{i}(\lambda,{\bf {\bf 0}})[\bar{v}(mv,\lambda_{2}')\slashed{\bar{n}}v(\hat{p}_{\bar{Q}},\lambda_{2})]
\\ \nonumber &
  +\frac{g^{2}C_{F}}{16m^{2}\sqrt{z(1-z)}}(\theta(z-\frac{1}{2})(1-z)+\theta(\frac{1}{2}-z)z)\left(K_{0}(\tau)-\frac{\tau}{2}K_{1}(\tau)\right)
\\ \nonumber &
\quad\quad\times\sum_{\lambda_{1}',\lambda_{2}'}[\bar{u}(\hat{p}_{Q},\lambda_{1})\slashed{\bar{n}}[\gamma_{\perp}^{i},\gamma_{\perp}^{j}]u(mv,\lambda_{1}')]\int\frac{\,d\lambda}{4\pi}\phi_{q\bar{q}}^{i}(\lambda,{\bf {\bf 0}})[\bar{v}(mv,\lambda_{2}')[\gamma_{\perp}^{j},\gamma_{\perp}^{i}]\slashed{\bar{n}}v(\hat{p}_{\bar{Q}},\lambda_{2})]
\\ \nonumber &
 +\frac{g^{2}C_{F}\left(z-\frac{1}{2}\right)r_{\perp}}{2m\sqrt{z(1-z)}}(\theta(z-\frac{1}{2})(1-z)+\theta(\frac{1}{2}-z)z)K_{1}(\tau)
\\ \nonumber &
\quad \quad \times \sum_{\lambda_{1}',\lambda_{2}'}[\bar{u}(\hat{p}_{Q},\lambda_{1})\gamma_{\perp}^{i}\slashed{\bar{n}}u(mv,\lambda_{1}')]\int\frac{\,d\lambda}{4\pi}\phi_{q\bar{q}}^{i}(\lambda,{\bf {\bf 0}})[\bar{v}(mv,\lambda_{2}')\gamma_{\perp}^{i}\slashed{\bar{n}}v(\hat{p}_{\bar{Q}},\lambda_{2})]\,.
\end{align}
Decomposing this into meson polarization states we get for the longitudinal polarization
\begin{multline}
\label{eq:Langavg}
\int\,d\theta_{r}  \left.\left\{ \Psi_{HQ}^{q\bar{q}}\right\} _{\lambda_{1}\lambda_{2}}^{3}(z,r_{\perp})\right|_{Fig.\ref{fig:kernelr-1}}  =\frac{4g^{2}C_{F}z(1-z)}{\left(z-\frac{1}{2}\right)^{2}}
\bigg\{K_{0}(\tau)
\\
+\left(\theta(z-\frac{1}{2})(1-z)+\theta(\frac{1}{2}-z)z\right)
\bigg[\frac{\left(z-\frac{1}{2}\right)^{2}-\frac{1}{2}}{z(1-z)}\left(K_{0}(\tau)-\frac{\tau}{2}K_{1}(\tau)\right)
-\frac{\left(z-\frac{1}{2}\right)^{2}}{2z(1-z)}\tau K_{1}(\tau)\bigg]\bigg\}
\left\{\int\frac{\,d\lambda}{4\pi}\phi_{q\bar{q}}^{3}(\lambda,{\bf {\bf 0}})\right\}_{\lambda_1,\lambda_2}\,,
 \end{multline}
while for the transverse case, we obtain
\begin{multline}
\int\,d\theta_{r}\left.\left\{ \Psi_{HQ}^{q\bar{q}}\right\} _{\lambda_{1}\lambda_{2}}^{i}(z,r_{\perp})\right|_{Fig.\ref{fig:kernelr-1}}
 =\frac{4g^{2}C_{F}z(1-z)}{\left(z-\frac{1}{2}\right)^{2}}
 \bigg\{ K_{0}(\tau)
 \\
 -\frac{(\theta(z-\frac{1}{2})(1-z)+\theta(\frac{1}{2}-z)z)}{2z(1-z)}
 \left(K_{0}(\tau)-\frac{\tau}{2}K_{1}(\tau)\right)
 \bigg\}
 \left\{\int\frac{\,d\lambda}{4\pi}\phi_{q\bar{q}}^{i}(\lambda,{\bf {\bf 0}})\right\}_{\lambda_1,\lambda_2}\,.
\end{multline}
It is also possible to write a compact expression in coordinate space for the polar angle average of the coefficients
$C_{q\bar{q}\gets q\bar{q}}^{0}(z,{\bf p}_{\perp};\lambda_{1},\lambda_{2};\lambda_{1}',\lambda_{2}')$.   For the case of longitudinal
polarization, the result is 
\begin{multline}
\int\,d\theta_{r}C_{q\bar{q}\gets q\bar{q}}^{0}(z,{\bf r}_{\perp};\lambda_{1},\lambda_{2};\lambda_{1}',\lambda_{2}')_{long}=
8\pi^{2}\delta(z-\frac{1}{2})(1+\delta Z)
\delta_{\lambda_{1}\lambda_{1}'}\delta_{\lambda_{2}\lambda_{2}'}
+\frac{4g^{2}C_{F}z(1-z) }{\left(z-\frac{1}{2}\right)^{2}}
\delta_{\lambda_{1},\lambda_{1}'}\delta_{\lambda_{2},\lambda_{2}'}
\Bigg\{ K_{0}(\tau)
\\
+\left(\theta(z-\frac{1}{2})(1-z)+\theta(\frac{1}{2}-z)z\right)\left[\frac{\left(z-\frac{1}{2}\right)^{2}-\frac{1}{2}}{z(1-z)}\left(K_{0}(\tau)-\frac{\tau}{2}K_{1}(\tau)\right)-\frac{\left(z-\frac{1}{2}\right)^{2}}{2z(1-z)}\tau K_{1}(\tau)\right]\Bigg\}
\,,
\label{eq:Cqqtoqqlong}
\end{multline}
and for the transverse polarization
\begin{multline}
\int\,d\theta_{r}C_{q\bar{q}\gets q\bar{q}}^{0}(z,{\bf r}_{\perp};\lambda_{1},\lambda_{2};\lambda_{1}',\lambda_{2}')_{trans}=8\pi^{2}\delta(z-\frac{1}{2})(1+\delta Z)\delta_{\lambda_{1}\lambda_{1}'}\delta_{\lambda_{2}\lambda_{2}'}
\\
+\frac{4g^{2}C_{F}z(1-z)\delta_{\lambda_{1},\lambda_{1}'}\delta_{\lambda_{2},\lambda_{2}'}}{\left(z-\frac{1}{2}\right)^{2}}\left[K_{0}(\tau)-\frac{(\theta(z-\frac{1}{2})(1-z)+\theta(\frac{1}{2}-z)z)}{2z(1-z)}\left(K_{0}(\tau)-\frac{\tau}{2}K_{1}(\tau)\right)\right]
\,.
\label{eq:Cqqtoqqtran}
\end{multline}

\begin{figure}
\centerline{\includegraphics[width=0.5\textwidth]{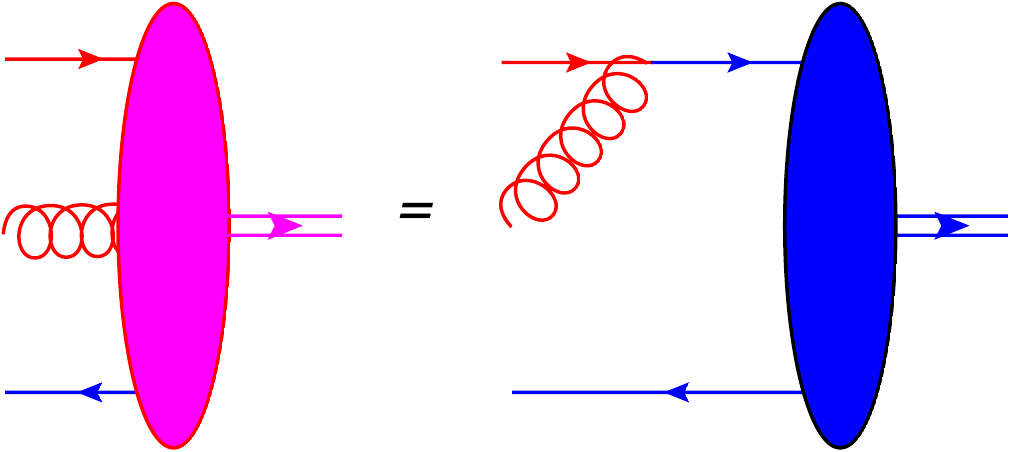}}

\caption{\label{fig:bethe3} LO contribution to the component of the wave function
with one relativistic heavy quark, one non-relativistic heavy antiquark
and a hard gluon.}
\end{figure}

Now let us move to the second contribution, where we have a relativistic
quark, a non-relativistic one and a hard gluon. This is represented
at leading order in the diagram of fig. \ref{fig:bethe3}. We need
to introduce some notation in order to write the result. The longitudinal
momentum fractions of the relativistic quark, the gluon and the non-relativistic
antiquark are respectively $z_{RQ}=\frac{1-x}{2}+\lambda$, $z_{G}=\frac{x}{2}$
and $z_{\bar{Q}}=\frac{\text{1}}{2}-\lambda$. From the nonrelativistic nature of the ``spectator'' antiquark it follows that $\lambda\ll1$. We shall consequently not write out the integration limits for $\lambda$, since it is understood to be dominated by a short interval around the origin.
The transverse momenta are respectively ${\bf P}_{\perp,RQ}=\frac{{\bf p}_{\perp}}{2}-{\bf q}_{\perp}$,
${\bf P}_{\perp,G}=\frac{{\bf p}_{\perp}}{2}+{\bf q}_{\perp}$ and
${\bf P}_{\perp,\bar{Q}}=-{\bf p}_{\perp}$, 
where again for the nonrelativistic antiquark ${\bf p}_{\perp}\ll m$.
Using this we get
\begin{multline}
\int\frac{\,d\lambda}{4\pi}\frac{\,d^{D-2}p_{\perp}}{(2\pi)^{D-2}}\left\{ \Psi_{HQ}^{q\bar{q}g}\right\} _{\lambda_{RQ},\lambda_{G},\lambda_{\bar{Q}}}^{i}(x,{\bf q}_{\perp};\lambda,{\bf p}_{\perp})
\\
=-g\frac{\sqrt{x(1-x)}}{\sqrt{2}\left(q_{\perp}^{2}+m^{2}x^{2}\right)}\sum_{\lambda_{Q}}\bar{u}(\hat{p}_{RQ},\lambda_{RQ})\slashed{\epsilon}^{*}(\hat{p}_{G},\lambda_{G})u(mv,\lambda_{Q})\int\frac{\,d\lambda}{4\pi}\phi_{q\bar{q}}^{i}(\lambda,{\bf {\bf 0}})\,,\label{eq:qbqgm-1}
\end{multline}
or equivalently
\begin{multline}
\int\frac{\,d\lambda}{4\pi}\frac{\,d^{D-2}p_{\perp}}{(2\pi)^{D-2}}\left\{ \Psi_{HQ}^{q\bar{q}g}\right\} _{\lambda_{RQ},\lambda_{G},\lambda_{\bar{Q}}}^{i}(x,{\bf q}_{\perp};\lambda,{\bf p}_{\perp})  =
\\
=-\frac{g\left(1-\frac{x}{2}\right)q_{\perp}^{i}{\epsilon_{\perp}^{*}}^{A,j}(\lambda_{G})}{\sqrt{x(1-x)}m(q_{\perp}^{2}+m^{2}x^{2})}\sum_{\lambda_{Q}}\bar{u}(\hat{p}_{RQ},\lambda_{RQ})T^{A}\left[\delta^{ij}-\frac{x}{4\left(1-\frac{x}{2}\right)}[\gamma_{\perp}^{i},\gamma_{\perp}^{j}]\right]\slashed{\bar{n}}u(mv,\lambda_{Q})\int\frac{\,d\lambda}{4\pi}\phi_{q\bar{q}}^{i}(\lambda,{\bf {\bf 0}})
\\
+\frac{g\sqrt{x(1-x)}}{2(1-x)}\frac{x}{q_{\perp}^{2}+m^{2}x^{2}}\sum_{\lambda_{Q}}\bar{u}(\hat{p}_{RQ},\lambda_{RQ})\slashed{\epsilon}_{\perp}^{*}(\lambda_{G})\slashed{\bar{n}}u(mv,\lambda_{Q})\int\frac{\,d\lambda}{4\pi}\phi_{q\bar{q}}^{i}(\lambda,{\bf {\bf 0}})\,.\label{eq:qbqgm}
\end{multline}
In order to write the result in coordinate space we define ${\bf r}_{\perp}$
as the Fourier conjugate of the nonrelativistic momentum ${\bf p}_{\perp}$ and ${\bf l}_{\perp}$ as the one of the relativistic one ${\bf q}_{\perp}$, leading to 
\begin{multline}
\int\frac{\,d\lambda}{4\pi}\left\{ \Psi_{HQ}^{q\bar{q}g}\right\} _{\lambda_{RQ},\lambda_{G},\lambda_{\bar{Q}}}^{i}(x,{\bf l}_{\perp};\lambda,{\bf r}_{\perp}={\bf 0})=
\\
-\frac{ig\left(1-\frac{x}{2}\right)xl_{\perp}^{i}{\epsilon_{\perp}^{*}}^{A,j}(\lambda_{G})}{2\pi\sqrt{x(1-x)}l_{\perp}}\left(\frac{mx}{2\pi l_{\perp}\mu^{2}}\right)^{\frac{D-4}{2}}K_{\frac{D-2}{2}}(mxl_{\perp})\sum_{\lambda_{Q}}\bar{u}(\hat{p}_{RQ},\lambda_{RQ})T^{A}
\\
\times
\left[\delta^{ij}-\frac{x}{4\left(1-\frac{x}{2}\right)}[\gamma_{\perp}^{i},\gamma_{\perp}^{j}]\right]\slashed{\bar{n}}u(mv,\lambda_{Q})\int\frac{\,d\lambda}{4\pi}\phi_{q\bar{q}}^{i}(\lambda,{\bf {\bf 0}})
\\
+\frac{g\sqrt{x(1-x)}x}{4\pi(1-x)}\left(\frac{mx}{2\pi l_{\perp}\mu^{2}}\right)^{\frac{D-4}{2}}K_{\frac{D-4}{2}}(mxl_{\perp})\sum_{\lambda_{Q}}\bar{u}(\hat{p}_{RQ},\lambda_{RQ})\slashed{\epsilon}_{\perp}^{*}(\lambda_{G})\slashed{\bar{n}}u(mv,\lambda_{Q})\int\frac{\,d\lambda}{4\pi}\phi_{q\bar{q}}^{i}(\lambda,{\bf {\bf 0}})\,.\label{eq:hqqbarqg}
\end{multline}
It is straightforward to obtain $C_{q\bar{q}g\gets q\bar{q}}^{0}$ from
the previous results as
\begin{multline}
C_{q\bar{q}g\gets q\bar{q}}^{0}(x,{\bf l}_{\perp};\lambda,{\bf r}_{\perp}={\bf 0};\lambda_1,\lambda_G,\lambda_2;\lambda_1',\lambda_2')=\\
-\frac{ig\left(1-\frac{x}{2}\right)xl_{\perp}^{i}{\epsilon_{\perp}^{*}}^{A,j}(\lambda_{G})}{2\pi\sqrt{x(1-x)}l_{\perp}}\delta_{\lambda_2,\lambda_2'}\left(\frac{mx}{2\pi l_{\perp}\mu^{2}}\right)^{\frac{D-4}{2}}K_{\frac{D-2}{2}}(mxl_{\perp})\sum_{\lambda_{Q}}\bar{u}(\hat{p}_{RQ},\lambda_1)T^{A}
\\
\times
\left[\delta^{ij}-\frac{x}{4\left(1-\frac{x}{2}\right)}[\gamma_{\perp}^{i},\gamma_{\perp}^{j}]\right]\slashed{\bar{n}}u(mv,\lambda_1')
\\
+\frac{g\sqrt{x(1-x)}x}{4\pi(1-x)}\delta_{\lambda_2,\lambda_2'}\left(\frac{mx}{2\pi l_{\perp}\mu^{2}}\right)^{\frac{D-4}{2}}K_{\frac{D-4}{2}}(mxl_{\perp})\sum_{\lambda_{Q}}\bar{u}(\hat{p}_{RQ},\lambda_1)\slashed{\epsilon}_{\perp}^{*}(\lambda_{G})\slashed{\bar{n}}u(mv,\lambda_1')\,.
\label{eq:46}
\end{multline}

A cross-check of the results of this section can be obtained by checking
that the normalization of the light-cone wave-function is not changed
by radiative corrections. To order $\alpha_{s}$ this is ensured if
the square of the diagram in fig. \ref{fig:bethese} cancels with
the square of the diagram in fig. \ref{fig:bethe3}. An easy way
to check this is to start from eq. (\ref{eq:qbqgm-1}), after doing
the square and performing the trace over gamma matrices one immediately
obtains an equation compatible with (3.28) in \cite{Mustaki:1990im},
which is equal to $-\delta Z$.

\section{Light-cone distribution amplitudes\label{sec:Light-cone-distribution-amplitud}}

In the cases in which the virtuality of the photon is much larger
than other dimensionful scales in the problem, in particular the mass of the heavy quark, one can use collinear factorization
\cite{Lepage:1980fj,Chernyak:1983ej}. The information about the bound
state in this formalism is encoded in the light-cone distribution
amplitude (DA). This coincides with the light-cone wave function in the
limit in which the transverse radius $r_{\perp}$ goes to zero. This
object was already studied in \cite{Ma:2006hc,Bell:2008er,Jia:2008ep}  using the non-relativistic
expansion (see also \cite{Ding:2015rkn}), here we provide an independent 
computation in light-cone gauge. 

Naively, one would expect that one can obtain the light-cone distribution
amplitude trivially making the limit $r_{\perp}\to0$ in the results
in the previous section. This is not so because taking the limit introduces
an ultraviolet divergence that needs to be regularized. This can be done
in dimensional regularization using the following prescription
\begin{equation}
\lim_{r_{\perp}\to0}f(z,{\bf r}_{\perp})=\int\frac{\,d^{D-2}p_{\perp}}{(2\pi)^{D-2}}f(z,{\bf p}_{\perp})\,.
\end{equation}
Applying this method to eq. (\ref{eq:psirel}) one obtains for the longitudinal
polarization
\begin{align}
\left.\left\{ \Psi_{HQ}^{q\bar{q}}\right\} _{\lambda_{1}\lambda_{2}}^{3}(z,r_{\perp}=0)\right|_{Fig.\ref{fig:kernelr-1}} & =-\frac{g^{2}C_{F}\sqrt{z(1-z)}}{2\pi m^{2}\left(z-\frac{1}{2}\right)^{2}}\left[\left(\frac{1}{D-4}+\frac{1}{2}\log\left(\frac{m^{2}\left(z-\frac{1}{2}\right)^{2}}{\pi\mu^{2}}\right)+\frac{\gamma_{E}}{2}\right)\times\right.\nonumber \\
 & \left.\times\left(1+(\theta(z-\frac{1}{2})(1-z)+\theta(\frac{1}{2}-z)z)\frac{(D-2)\left(z-\frac{1}{2}\right)^{2}-1}{2z(1-z)}\right)+\right.\nonumber \\
 & \left.+\frac{(\theta(z-\frac{1}{2})(1-z)+\theta(\frac{1}{2}-z)z)}{4z(1-z)}\left(2(D-2)\left(z-\frac{1}{2}\right)^{2}-1\right)\right]\times\nonumber \\
 & \times\sum_{\lambda_{1}',\lambda_{2}'}[\bar{u}(\hat{p}_{Q},\lambda_{1})\slashed{\bar{n}}u(mv,\lambda_{1}')]\int\frac{\,d\lambda}{4\pi}\phi_{q\bar{q}}^{3}(\lambda,{\bf {\bf 0}})[\bar{v}(mv,\lambda_{2}')\slashed{\bar{n}}v(\hat{p}_{\bar{Q}},\lambda_{2})]\,,
\end{align}
and for the transverse one
\begin{align}
\left.\left\{ \Psi_{HQ}^{q\bar{q}}\right\} _{\lambda_{1}\lambda_{2}}^{i}(z,r_{\perp}=0)\right|_{Fig.\ref{fig:kernelr-1}} & =-\frac{g^{2}C_{F}\sqrt{z(1-z)}}{2\pi m^{2}\left(z-\frac{1}{2}\right)^{2}}\left[\left(\frac{1}{D-4}+\frac{1}{2}\log\left(\frac{m^{2}\left(z-\frac{1}{2}\right)^{2}}{\pi\mu^{2}}\right)+\frac{\gamma_{E}}{2}\right)\times\right.\nonumber \\
 & \left.\left(1-\frac{(\theta(z-\frac{1}{2})(1-z)+\theta(\frac{1}{2}-z)z)}{2z(1-z)}\right)-\frac{(\theta(z-\frac{1}{2})(1-z)+\theta(\frac{1}{2}-z)z)}{4z(1-z)}\right]\times\nonumber \\
 & \times\sum_{\lambda_{1}',\lambda_{2}'}[\bar{u}(\hat{p}_{Q},\lambda_{1})\slashed{\bar{n}}u(mv,\lambda_{1}')]\int\frac{\,d\lambda}{4\pi}\phi_{q\bar{q}}^{i}(\lambda,{\bf {\bf 0}})[\bar{v}(mv,\lambda_{2}')\slashed{\bar{n}}v(\hat{p}_{\bar{Q}},\lambda_{2})]\,.
\end{align}

We now define the light-cone distribution amplitude as
\begin{equation}
D^{i}(z)=\left\{ \Psi_{HQ}^{q\bar{q}}\right\} _{\lambda_{1}\lambda_{2}}^{i}(z,r_{\perp}=0)\,,
\end{equation}
Using the previous results, the light-cone distribution amplitude
$D(z)$ can be written as
\begin{equation}
D^{i}(z)=4\pi(1+\delta Z)\delta\left(z-\frac{1}{2}\right)\int\frac{\,d\lambda}{4\pi}\phi_{q\bar{q}}^{i}(\lambda,{\bf {\bf 0}})+\left.\left\{ \Psi_{HQ}^{q\bar{q}}\right\} _{\lambda_{1}\lambda_{2}}^{i}(z,r_{\perp}=0)\right|_{Fig.\ref{fig:kernelr-1}}\,.
\end{equation}
For the longitudinal polarization, this is 
\begin{align}
D^{3}(z)= & 4\pi(1+\delta Z)\delta\left(z-\frac{1}{2}\right)\int\frac{\,d\lambda}{4\pi}\phi_{q\bar{q}}^{3}(\lambda,{\bf {\bf 0}})
\nonumber \\
 & -\frac{2g^{2}C_{F}z(1-z)}{\pi\left(z-\frac{1}{2}\right)^{2}}\left[\left(\frac{1}{D-4}+\frac{1}{2}\log\left(\frac{m^{2}\left(z-\frac{1}{2}\right)^{2}}{\pi\mu^{2}}\right)+\frac{\gamma_{E}}{2}\right) \right.
 \nonumber \\
 & \quad\quad \left.\times\left(1+(\theta(z-\frac{1}{2})(1-z)+\theta(\frac{1}{2}-z)z)\frac{(D-2)\left(z-\frac{1}{2}\right)^{2}-1}{2z(1-z)}\right)\right.
 \nonumber \\
 &  \left.+\frac{(\theta(z-\frac{1}{2})(1-z)+\theta(\frac{1}{2}-z)z)}{4z(1-z)}\left(2(D-2)\left(z-\frac{1}{2}\right)^{2}-1\right)\right]\int\frac{\,d\lambda}{4\pi}\phi_{q\bar{q}}^{3}(\lambda,{\bf {\bf 0}})\,,\label{eq:D3}
\end{align}
while for the transverse one the DA is
\begin{multline}
D^{i}(z)=  4\pi(1+\delta Z)\delta\left(z-\frac{1}{2}\right)\int\frac{\,d\lambda}{4\pi}\phi_{q\bar{q}}^{i}(\lambda,{\bf {\bf 0}})
\\
 -\frac{2g^{2}C_{F}z(1-z)}{\pi\left(z-\frac{1}{2}\right)^{2}}
 \bigg[
 \left(\frac{1}{D-4}+\frac{1}{2}\log\left(\frac{m^{2}\left(z-\frac{1}{2}\right)^{2}}{\pi\mu^{2}}\right)+\frac{\gamma_{E}}{2}\right)
 \left(1-\frac{(\theta(z-\frac{1}{2})(1-z)+\theta(\frac{1}{2}-z)z)}{2z(1-z)}\right)
   \\ 
 -\frac{(\theta(z-\frac{1}{2})(1-z)+\theta(\frac{1}{2}-z)z)}{4z(1-z)}\bigg]
 \int\frac{\,d\lambda}{4\pi}\phi_{q\bar{q}}^{i}(\lambda,{\bf {\bf 0}})\,.
\label{eq:Di}
 \end{multline}
The first lines in \eqs\nr{eq:D3} and~\nr{eq:Di} are the leading order bare amplitude (apart from the $\delta Z$). The NLO corrections involve a divergence proportional to $\sim 1/(D-4)$ from the transverse integrals, and the associated logarithm of the regularization scale $\mu$. In the usual fashion these divergences must be absorbed into the bare distribution, leaving the renormalized distribution dependent on the scale $\mu$. This dependence is related to the ERBL evolution equation
\cite{Lepage:1980fj,Chernyak:1983ej}:
\begin{equation}
\frac{\partial D^{i}(z)}{\partial\log\mu^{2}}=\frac{\alpha_{s}C_{F}}{2\pi}\int_{0}^{1}\,dz'K_{L,T}(z,z')D^{i}(z')\,.
\label{eq:erbl}
\end{equation}

To check that we recover this equation in the standard form  we can take the expressions for the kernels in \eq\nr{eq:erbl} from the literature and compute the r.h.s. The comparison of this result with the coefficient of the logarithm of $\mu$ in  \eqs\nr{eq:D3} and~\nr{eq:Di} then provides a check of our calculation of the wave function.

The ERBL kernel for the longitudinal polarization is given by
\begin{equation}
K_{L}(z,z')=\theta(z-z')\frac{1-z}{1-z'}\left(1+\left[\frac{1}{z-z'}\right]_{+}\right)+\theta(z'-z)\frac{z}{z'}\left(1+\left[\frac{1}{z'-z}\right]_{+}\right)+\frac{3}{2}\delta(z-z')\,,
\end{equation}
while for transverse one it is
\begin{equation}
K_{T}(z,z')=\theta(z-z')\frac{1-z}{1-z'}\left[\frac{1}{z-z'}\right]_{+}+\theta(z'-z)\frac{z}{z'}\left[\frac{1}{z'-z}\right]_{+}+\frac{3}{2}\delta(z-z')\,.
\end{equation}
The $[]_{+}$ notation means that 
\begin{equation}
\int\,dzf(z)\left[\frac{1}{z-z_{0}}\right]_{+}=\int\,dz\frac{f(z)-f(z_{0})}{z-z_{0}}\,.
\end{equation}
Note that the $[]_{+}$prescription can be implemented with a cut-off,
as we did when computing $\delta Z$. For example,
\begin{equation}
\int\,dz\theta\left(z-\frac{1}{2}\right)f(z)\left[\frac{1}{z-\frac{1}{2}}\right]_{+}=\int_{\frac{1+x_{0}}{2}}^{1}\,dz\frac{f(z)}{z-\frac{1}{2}}+f\left(\frac{1}{2}\right)\log x_{0}\,.
\end{equation}

We can then check that the light-cone distribution amplitudes that
we have computed fulfil the ERBL equation by computing explicitly
$\frac{\partial D^{i}(z)}{\partial\log\mu^{2}}$. In the case of longitudinal
polarization
\begin{align}
\frac{\partial D^{3}(z)}{\partial\log\mu^{2}}= & 2\alpha_{s}C_{F}\left[2\log x_{0}+\frac{3}{2}\right]\delta\left(z-\frac{1}{2}\right)\int\frac{\,d\lambda}{4\pi}\phi_{q\bar{q}}^{3}(\lambda,{\bf {\bf 0}})\nonumber \\
 & +4\alpha_{s}C_{F}\left[\theta\left(z-\frac{1}{2}\right)(1-z)\left(1+\frac{1}{z-\frac{1}{2}}\right)+\theta\left(\frac{1}{2}-z\right)z\left(1+\frac{1}{\frac{1}{2}-z}\right)\right]\int\frac{\,d\lambda}{4\pi}\phi_{q\bar{q}}^{3}(\lambda,{\bf {\bf 0}})\,,
\end{align}
while for the transverse one 
\begin{align}
\frac{\partial D^{i}(z)}{\partial\log\mu^{2}}= & 2\alpha_{s}C_{F}\left[2\log x_{0}+\frac{3}{2}\right]\delta\left(z-\frac{1}{2}\right)\int\frac{\,d\lambda}{4\pi}\phi_{q\bar{q}}^{i}(\lambda,{\bf {\bf 0}}) \nonumber \\
 & +4\alpha_{s}C_{F}\left[\theta\left(z-\frac{1}{2}\right)\frac{1-z}{z-\frac{1}{2}}+\theta\left(\frac{1}{2}-z\right)\frac{z}{\frac{1}{2}-z}\right]\int\frac{\,d\lambda}{4\pi}\phi_{q\bar{q}}^{i}(\lambda,{\bf {\bf 0}})\,.
\end{align}
Thus we have verified that we recover a standard ERBL evolution of the light cone wave function. This is a check of  our NLO wavefunction  for the quark-antiquark Fock state in the short distance limit that is probed in a high virtuality (large $Q^2$) process. We emphasize that the quark momentum scale (or coordinate separation) in the whole wave function \nr{eq:qbarqmom}  is controlled by the quark mass, and this full result can be used more generally than just in the large $Q^2$ limit. 

\section{Radiative corrections to s-wave quarkonium decay\label{sec:Radiative-corrections-to}}

The decay width of quarkonia to leptons offers an important constraint on the nonrelativistic wavefunction. In this section we will see how the decay width is related to the light cone wavefunction. We will see that our regularization involving a cutoff in the $p^+$ momentum modifies the relation between the nonrelativistic wavefunction and the decay width compared to a usual rotationally invariant nonrelativistic potential model calculation.

\begin{figure}
\includegraphics{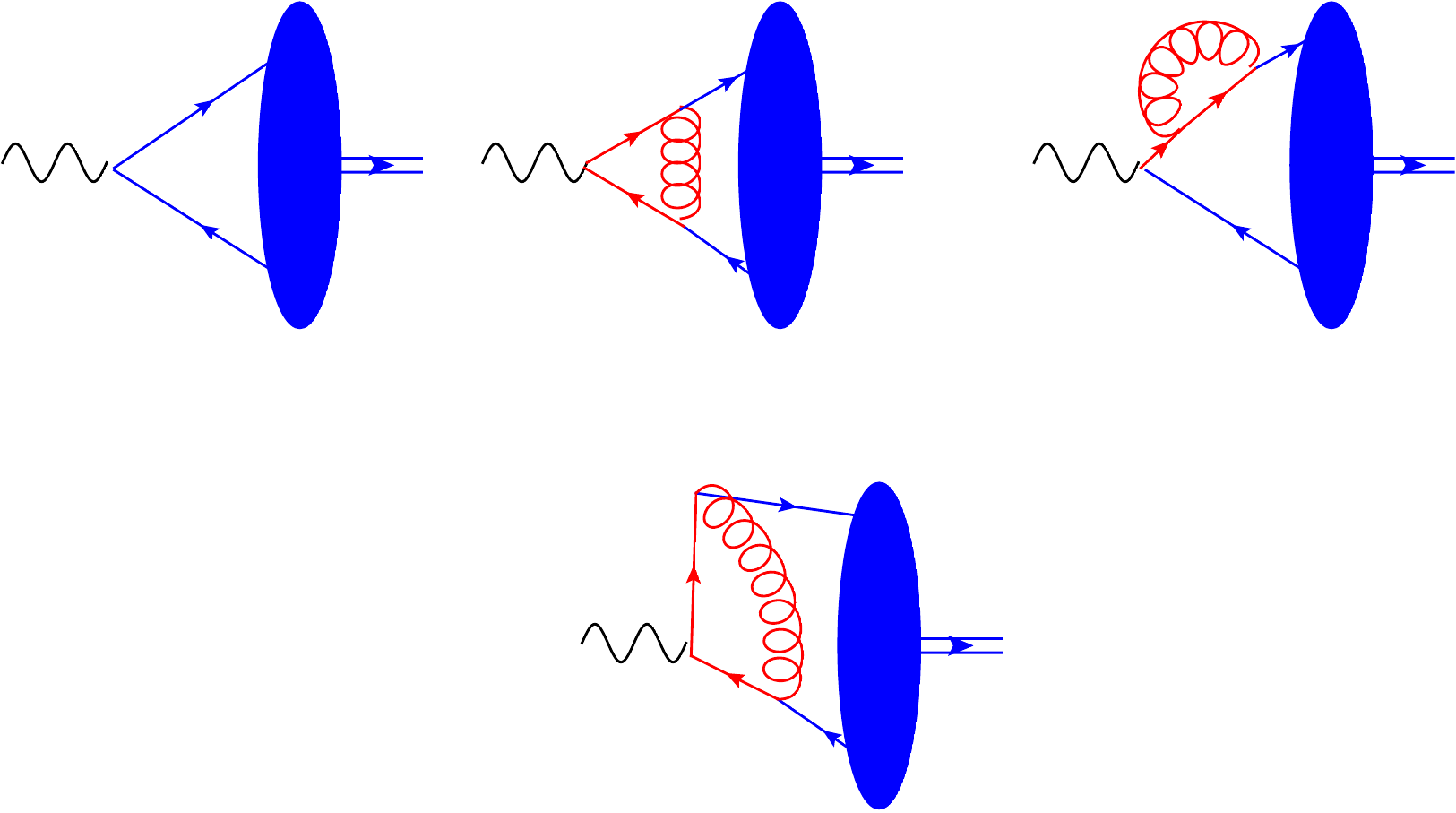}

\caption{Diagrams contributing to the overlap between the vector meson state and the electromagnetic current (photon). The colour
code is the same as in previous diagrams. In the diagram in the second
line, the vertical quark line represents an instantaneous interaction
in light-cone perturbation theory. This diagram vanishes for longitudinal
polarization.\label{fig:Diagrams-contributing-to}}

\end{figure}

The decay of quarkonium in a S-wave state into leptons is expressed in terms of the meson decay constant, which is defined as the matrix element of the electromagnetic current operator with the vector meson state \cite{Frankfurt:1997fj}. The current operator is local. As a consequence, at leading order, the decay constant is related to the value of the wave function at the origin. This is true  both for the light cone wave function and the nonrelativistic wavefunction. Beyond leading order, the relation between the decay constant and the nonrelativistic wavefunction receives a NLO correction~\cite{Barbieri:1975ki}. In light cone perturbation theory, on the other hand, the equivalent diagrams become corrections to the light cone wavefunction itself, not to the relation with the decay constant. Equating the decay constants in terms of the light cone and nonrelativistic wavefunctions yields the equation
\begin{equation}
\int_{0}^{1}\,dz\sum_{n}\Psi_{HQ}^{n}(z,{\bf 0}_{\perp})=\left(1-\frac{2\alpha_{s}C_{F}}{\pi}\right)\int d\lambda\phi(\lambda,{\bf 0})\,,
\end{equation}
where the sum in $n$ is over all the Fock state components that contribute
to the wave-function at the origin. In the case of longitudinal polarization,
the only component at NLO is $n=q\bar{q}$. For the transverse polarization,
there is also a contribution where the overlap between the electromagnetic current operator and the vector meson state receives a correction from the $n=q\bar{qg}$ Fock state, mediated by an instantanous interaction. The different one-loop corrections are 
shown diagrammatically in fig. (\ref{fig:Diagrams-contributing-to}). 

In the following we shall concentrate, for simplicity, on the longitudinal polarization state.
 In this case we have, by construction
\begin{equation}
\int_{0}^{1} dz\Psi_{HQ}^{q\bar{q}}(z,{\bf 0}_{\perp})=\int_{0}^{1} d z D(z)\,.
\end{equation}
Therefore, we can do the computation using the light-cone distribution
amplitude that we computed in the previous section. It is useful to
note that
\begin{equation}
\int_{0}^{1} dzK_{L}(z,1/2)=0\,.
\end{equation}
Using this on eq. (\ref{eq:D3}) we get 
\begin{equation}
\int_{0}^{1}\,dz\Psi_{HQ}^{q\bar{q}}(z,{\bf 0}_{\perp})=\left(1+\frac{\alpha_{s}C_{F}}{\pi}\left(\frac{1}{x_{0}}-2\right)\right)\int d\lambda\phi(\lambda,{\bf 0})\,.
\label{eq:lcvsnonrel}
\end{equation}
We see that, apart from a finite piece that coincides with the result
in \cite{Barbieri:1975ki}, we also get a divergent term. Its origin
is the Coulomb singularity, which would not be present at this order in dimensional
regularization. 

Degrees of freedom with transverse momentum much smaller than $m$
and $z-\frac{1}{2}$ close to zero should be included only in $\phi$.
However, when we computed eq. (\ref{eq:longwf}) we did not include
any mechanism to subtract them. Looking at eq. (\ref{eq:psirel})
we can see that, the only scale that affects the transverse momentum
when computing $\Psi_{HQ}^{q\bar{q}}(z,{\bf 0}_{\perp})$ is $2m\left(z-\frac{1}{2}\right)$.
Then, because we are using dimensional regularization in the $p_{\perp}$
integration, if we choose a cut-off $x_{0}$ much bigger than the
non-relativistic velocity around the centre of mass $v$ we can make
sure of only taking into account relativistic degrees of freedom.
However, $\int d\lambda\phi(\lambda,{\bf 0})$    will also
depend on $x_{0}$. The condition that must be fulfilled is that the decay constant is independent of the cutoff $x_0$:
\begin{equation}
\frac{d}{dx_{0}}\int_{0}^{1}\,dz\Psi_{HQ}^{q\bar{q}}(z,{\bf 0}_{\perp})
=-\frac{\alpha_{s}C_{F}}{\pi x_{0}^{2}}\int d\lambda\phi(\lambda,{\bf 0})
+\frac{d}{dx_{0}}\int d\lambda\phi(\lambda,{\bf 0})=0\,.\label{eq:Coulombsin}
\end{equation}

In order to compute how $\int d\lambda\phi(\lambda,{\bf 0})$
depends on the cut-off we can make use of Bethe-Salpeter equation
represented in \fig\ref{fig:bethenr}, following the lines of the
discussion in Appendix~\ref{sec:Relativistic-correction-to}. Note
that from the non-relativistic point of view, $z\sim x_{0}$ or $1-z\sim x_0$
represent the ultraviolet since the quark relative longitudinal momentum is large. 
Therefore, we can substitute the kernel
by a one gluon exchange between the quark and the antiquark and the
non-relativistic part of the wave-function renormalization. Let us
begin with the computation of the one gluon exchange part of the r.h.s.
of the Bethe-Salpeter equation of \fig\ref{fig:bethenr}, that we name $\Psi_{sub}^{q\bar{q}}$,
\begin{equation}
\left\{ \Psi_{sub}^{q\bar{q}}\right\} ^{i}(z,p_{\perp})=\frac{2g^{2}C_{F}}{\left(z-\frac{1}{2}\right)^{2}\left(p_{\perp}^{2}+4m^{2}\left(z-\frac{1}{2}\right)^{2}\right)}\left(\left|z-\frac{1}{2}\right|+\frac{2m^{2}\left(z-\frac{1}{2}\right)^{2}\left(1-2\left|z-\frac{1}{2}\right|\right)}{p_{\perp}^{2}+4m^{2}\left(z-\frac{1}{2}\right)^{2}}\right)\int\frac{\,d\lambda}{4\pi}\phi_{q\bar{q}}^{i}(\lambda,{\bf {\bf 0}})\,.
\end{equation}
From this expression we can obtain,
\begin{align}
\left\{ \Psi_{sub}^{q\bar{q}}\right\} ^{i}(z,r_{\perp} & =0)=\frac{g^{2}C_{F}\left(1-2\left|z-\frac{1}{2}\right|\right)}{4\pi\left(z-\frac{1}{2}\right)^{2}}\int\frac{\,d\lambda}{4\pi}\phi_{q\bar{q}}^{i}(\lambda,{\bf {\bf 0}})\nonumber \\
 & -\frac{g^{2}C_{F}}{\pi\left|z-\frac{1}{2}\right|}\left(\frac{1}{D-4}+\frac{1}{2}\log\left(\frac{m^{2}\left(z-\frac{1}{2}\right)^{2}}{\pi\mu^{2}}\right)+\frac{\gamma_{E}}{2}\right)\int\frac{\,d\lambda}{4\pi}\phi_{q\bar{q}}^{i}(\lambda,{\bf {\bf 0}})\,,
\end{align}
and therefore
\begin{align}
\frac{d}{dx_{0}}\int\,dz\left\{ \Psi_{sub}^{q\bar{q}}\right\} ^{i}(z,x_{\perp} & =0)=4\alpha_{s}C_{F}\left(\frac{1}{x_{0}^{2}}-\frac{1}{x_{0}}-\frac{2}{x_{0}}\log x_{0}\right)\int\frac{\,d\lambda}{4\pi}\phi_{q\bar{q}}^{i}(\lambda,{\bf {\bf 0}})\nonumber \\
 & -\frac{8\alpha_{s}C_{F}}{ x_{0}}\left(\frac{1}{D-4}+\frac{1}{2}\log\left(\frac{m^{2}}{4\pi\mu^{2}}\right)+\frac{\gamma_{E}}{2}\right)\int\frac{\,d\lambda}{4\pi}\phi_{q\bar{q}}^{i}(\lambda,{\bf {\bf 0}})\,.
\end{align}

The non-relativistic part of the wave-function renormalization is
computed in Appendix~\ref{sec:The-wave-function-renormalizatio}.
Here we use that
\begin{align}
\frac{d}{dx_{0}}\delta Z_{NR} & =\frac{\alpha_{s}C_{F}}{\pi x_{0}}\left(1+2\log x_{0}\right)\nonumber \\
 & +\frac{2\alpha_{s}C_{F}}{\pi x_{0}}\left(\frac{1}{D-4}+\frac{1}{2}\log\left(\frac{m^{2}}{4\pi\mu^{2}}\right)+\frac{\gamma_{E}}{2}\right)\,.
\end{align}
Putting the two one-loop corrections  pieces together we get
\begin{equation}
\frac{d}{dx_{0}}\int\,d\lambda\phi_{q\bar{q}}^{i}(\lambda,{\bf {\bf 0}})=\frac{d}{dx_{0}}\delta Z_{NR}\int\,d\lambda\phi_{q\bar{q}}^{i}(\lambda,{\bf {\bf 0}})+\frac{d}{dx_{0}}\int\,dz\left\{ \Psi_{sub}^{q\bar{q}}\right\} ^{i}(z,x_{\perp}=0)=\frac{\alpha_{s}C_{F}}{\pi x_{0}^{2}}\int\,d\lambda\phi_{q\bar{q}}^{i}(\lambda,{\bf {\bf 0}})\,.
\end{equation}

We can then see that eq. (\ref{eq:Coulombsin}) is fulfilled and the dependence on the cutoff $x_0$ cancels. Our regularization scheme with a cutoff $x_0$ is not the same as conventional rotationally invariant dimensional regularization. Cutoff-dependent, not directly measurable, quantities do not need to be the same in both schemes. The nonrelativistic wavefunction at the origin is such a quantity; and indeed the relation between our wavefunction  and the dimensionally reduced one of e.g.~\cite{Barbieri:1975ki} is 
\begin{equation}
\int\,d\lambda\phi_{q\bar{q}}^{i}(\lambda,{\bf {\bf 0}})=\left(1-\frac{\alpha_{s}C_{F}}{x_{0}}\right)\int\,d\lambda\left\{ \phi_{q\bar{q}}^{i}(\lambda,{\bf {\bf 0}})\right\} _{DR}\,,
\end{equation}
where the subscript DR means that this would be the result that we would have obtained if we had regulated all divergences using dimensional regularization. Then the light cone wave function at the origin, i.e. the vector meson decay constant, is related to the regularized nonrelativistic wavefunction in the natural way
\begin{equation}
\int_{0}^{1}\,dz\Psi_{HQ}^{q\bar{q}}(z,{\bf 0}_{\perp})=\left(1-\frac{2\alpha_{s}C_{F}}{\pi}\right)\int_{0}^{1}\,d\lambda\left\{ \phi_{q\bar{q}}^{i}(\lambda,{\bf {\bf 0}})\right\} _{DR}\,.
\end{equation}

In conclusion, we recover the result in \cite{Barbieri:1975ki} after
taking into account the different renormalization prescription. To
our knowledge, this is the first computation of this relation (which
can be directly related with the decay into leptons) done in light-cone
perturbation theory.

Finally, let us discuss a more phenomenological explanation. We can
use the experimental value of the decay into electrons to obtain the
wave function at the origin as a function of $x_{0}$
\begin{equation}
\sum_{h,\bar{h}}\left|\int\frac{\,d\lambda}{4\pi}\phi^{q\bar{q}}\right|^{2}=\frac{3\pi m\Gamma_{\Psi\to e^{+}e^{-}}}{N_{c}e_{f}^{2}e^{4}}\left(1+\frac{2\alpha_{s}C_{F}}{\pi}\left(\frac{1}{x_{0}}-2\right)\right)+\mathcal{O}(\alpha_{s}^{2})\,.
\end{equation}

The formalism is consistent if we use the previous formula to compute
the relation of another observable with the decay width into electrons
and we get a result that does not depend on $x_{0}$. In the following
section we are going to see that this is indeed the case. 

\section{Exclusive quarkonium production\label{sec:Exclusive-quarkonium-production}}

\subsection{Leading order}

In this section, we study the cancellation of divergences in photoproduction
of quarkonium with longitudinal polarization at NLO. In particular, we will show that the dependence on the dimensional regularization scale $\mu$ cancels, and that the dependence on the longitudinal momentum cutoff $x_0$ can be factorized into the B-JIMWLK-evolution of the target.  To simplify the calculation we assume that $Q\gg m$. This assumption does not modify ultraviolet
or collinear divergences and has the advantage that we can use the
same photon wave-function that was used when studying massless quarks. Thus our results can to some extent  be seen as a reformulation of the result obtained for light vector mesons in a somewhat different language in Ref.~\cite{Boussarie:2016bkq}.

The LO result is given by eq. (\ref{eq:lo}). The LO $q\bar{q}$ component
of the photon wave-function is \cite{Kowalski:2006hc}:
\begin{equation}
\Psi_{\gamma}(z,r_{\perp})=ee_{f}\sqrt{N_{c}}\delta_{h,-\bar{h}}2Qz(1-z)
\frac{K_{0}\left(\sqrt{z(1-z)}Qr_{\perp}\right)}{2\pi}\,,
\end{equation}
where $e_{f}$ is the charge of the corresponding quark. Regarding
the cross-section $\sigma_{q\bar{q}}$, its specific form is model
dependent and in general phenomenologically realistic models do not allow for an analytical treatment. Instead, we will use here the  small $r_{\perp}$ (dilute) limit
\begin{equation}
\sigma_{q\bar{q}}(z,r_{\perp})\sim\frac{r_{\perp}^{2}\sigma_{0}Q_{s}^{2}(z)}{4}\,.
\end{equation}
In this paper, we are going to discuss in general the structure of
the divergences, in order to check the consistency of the method,
and we are going to apply the dilute limit to get analytic results
when possible.

Let us remind the reader of the LO result, which was computed in \cite{Brodsky:1994kf}.
This can be obtained by writing the LO photon wave-function in eq.
\nr{eq:lo} 
\begin{equation}
\left.\frac{d\sigma_{T,L}^{\gamma^{*}+N\to HQ+N}}{dt}\right|_{LO}=\frac{e^{2}e_{f}^{2}Q^{2}N_{c}}{256\pi^{3}}
\left|\sum_{h,\bar{h}}\int\frac{\,d\lambda}{4\pi}\phi^{q\bar{q}}\right|^{2}
\left|\int\,d^{2}r_{\perp}K_{0}\left(\frac{Qr_{\perp}}{2}\right)\sigma_{q\bar{q}}\left(\frac{1}{2},r_{\perp}\right)\right|^{2}\,.
\end{equation}
In the dilute limit, this becomes
\begin{equation}
\left.\frac{d\sigma_{T,L}^{\gamma^{*}+N\to HQ+N}}{dt}\right|_{LO,dilute}=\frac{4e^{2}e_{f}^{2}\sigma_{0}^{2}Q_{s}^{4}N_{c}}{\pi Q^{6}}
\left|\sum_{h,\bar{h}}\int\frac{\,d\lambda}{4\pi}\phi^{q\bar{q}}\right|^{2}\,.
\end{equation}
Using the relation between $Q_{s}$ and the gluon density from \cite{Kowalski:2006hc}
\begin{equation}
\sigma_{0}Q_{s}^{2}=\frac{4\pi^{2}}{N_{c}}\alpha_{s}(\mu^{2})xg(x,\mu^{2})\,,
\end{equation}
and the relation between quarkonium's wave-function at the origin
and its decay into electrons
\begin{equation}
\left.\Gamma_{\Psi\to e^{+}e^{-}}\right|_{LO}=\frac{N_{c}e_{f}^{2}e^{4}}{3\pi m}\sum_{h,\bar{h}}\left|\int\frac{\,d\lambda}{4\pi}\phi^{q\bar{q}}\right|^{2}\,,
\end{equation}
we get
\begin{equation}
\left.\frac{d\sigma_{T,L}^{\gamma^{*}+N\to HQ+N}}{dt}\right|_{LO,dilute}=\frac{96\pi^{3}m\alpha_{s}^{2}(\mu^{2})(xg(x,\mu^{2}))^{2}\Gamma_{\Psi\to e^{+}e^{-}}}{N_{c}^{2}\alpha Q^{6}}\,,
\end{equation}
which  agrees with the classic result from \cite{Ryskin:1992ui} (and is four times the result found in \cite{Brodsky:1994kf}). Now we  move on to discuss the different corrections that this process gets at NLO, focusing on the longitudinal polarization state at $Q\gg m$ where we can use results from the literature for massless quarks. At NLO, the production process gets contributions from both the $q\bar{q}$ and the $q\bar{q}g$  Fock states. In addition, we have to consider separately the corrections to the vector meson and the photon light cone wave functions, although in the large $Q^2$ limit they behave in a similar way.

\subsection{NLO corrections to the $q\bar{q}$ component of the photon wave-function}

We take the  correction  to the photon wave function from the recent computation~\cite{Beuf:2017bpd}
\begin{equation}
\left.\Psi_{\gamma}(z,r_{\perp})\right|_{NLO}=\left.\Psi_{\gamma}(z,r_{\perp})\right|_{LO}\left(1+\delta Z_{\gamma}(z,r_{\perp})\right)\,.
\end{equation}
In our case, we are going to need the expression for $z=\frac{1}{2}$
\begin{equation}
\delta Z_{\gamma}\left(\frac{1}{2},r_{\perp}\right)=-\frac{2C_{F}\alpha_{s}}{\pi}\left(\left(\log x_{0}+\frac{3}{4}\right)\left(\frac{1}{D-4}-\frac{\gamma_{E}}{2}-\frac{1}{2}\log(\pi\mu^{2}r_{\perp}^{2})\right)+\frac{\pi^{2}}{24}-\frac{3}{4}\right)\,.
\end{equation}

The effect of this correction on quarkonium production is the following
\begin{align}
\left.\frac{d\sigma_{T,L}^{\gamma^{*}+N\to HQ+N}}{dt}\right|_{NLO,\gamma} & =\frac{e^{2}e_{f}^{2}Q^{2}N_{c}}{128\pi^{3}}\left|\sum_{h,\bar{h}}\int\frac{\,d\lambda}{4\pi}\phi^{q\bar{q}}\right|^{2}\left(\int\,d^{2}r_{\perp}\delta Z_{\gamma}(\frac{1}{2},r_{\perp})K_{0}\left(\frac{Qr_{\perp}}{2}\right)\sigma_{q\bar{q}}\left(\frac{1}{2},r_{\perp}\right)\right)\times\nonumber \\
 & \times\left(\int\,d^{2}r'_{\perp}K_{0}\left(\frac{Qr'_{\perp}}{2}\right)\sigma_{q\bar{q}}\left(\frac{1}{2},r'_{\perp}\right)\right)^{*}\,.\label{eq:NLOA}
\end{align}
In the dilute limit, this is
\begin{align}
 & \left.\frac{d\sigma_{T,L}^{\gamma^{*}+N\to HQ+N}}{dt}\right|_{NLO,\gamma,dilute}=\nonumber \\
 & -\frac{16C_{F}\alpha_{s}e^{2}e_{f}^{2}\sigma_{0}^{2}Q_{s}^{4}N_{c}}{\pi^{2}Q^{6}}\left(\left(\log x_{0}+\frac{3}{4}\right)\left(\frac{1}{D-4}+\frac{\gamma_{E}}{2}+\frac{1}{2}\log\left(\frac{Q^{2}}{4\pi\mu^{2}}\right)-1-\log2\right)+\frac{\pi^{2}}{24}-\frac{3}{4}\right)\left|\sum_{h,\bar{h}}\int\frac{\,d\lambda}{4\pi}\phi^{q\bar{q}}\right|^{2}\,.\label{eq:A,dilute}
\end{align}

\subsection{NLO corrections to the $q\bar{q}$ component of quarkonium wave-function. }

In this subsection, we will compute one-loop corrections to the   $q\bar{q}$ component of quarkonium wave-function in the hard scattering limit. We will first study the divergence structure in the general case,  and then  perform the computation
in the dilute limit for the dipole-target scattering amplitude. Using \eq\nr{eq:Cexpand} the contribution that we are studying in this subsection is 
\begin{equation}
\left.\frac{d\sigma_{T,L}^{\gamma^{*}+N\to HQ+N}}{dt}\right|_{NLO,HQ}=\frac{\left|\int\frac{\,d\lambda}{4\pi}\phi^{q\bar{q}}\right|^{2}}{8\pi}\left(\int\,d^{2}r_{\perp}\int_{0}^{\text{1}}\,\frac{dz}{4\pi}\left(C_{q\bar{q}\gets q\bar{q}}^{0}\right)^{*}\left(\Psi_{\gamma^{*}}\right)_{L}\sigma_{q\bar{q}}\right)\left(\int\,d^{2}r'_{\perp}\left(\Psi_{\gamma^{*}}\right)_{L}\sigma_{q\bar{q}}\right)^{*}\,.
\end{equation}

First, we look at the dependence on $\mu$. For this, we note that
$\sigma_{q\bar{q}}$ goes like $r_{\perp}^{2}$ in the small $r_{\perp}$
region while both $C_{q\bar{q}\gets q\bar{q}}^{0}$ and $\left(\Psi_{\gamma^{*}}\right)_{L}$
go like $\log r_{\perp}$. On the other hand, for large $r_{\perp}$
both wave-functions are exponentially suppressed. Therefore we conclude
that the only dependence on $\mu$ comes from the wave-function
renormalization of the heavy quarks. Therefore
\begin{equation}
\frac{d}{d\mu}\left.\frac{d\sigma_{T,L}^{\gamma^{*}+N\to HQ+N}}{dt}\right|_{NLO,HQ}=\frac{e^{2}e_{f}^{2}Q^{2}N_{c}}{128\pi^{3}}\frac{d\delta Z}{d\mu}\left|\sum_{h,\bar{h}}\int\frac{\,d\lambda}{4\pi}\phi^{q\bar{q}}\right|^{2}\left|\int\,d^{2}r_{\perp}K_{0}\left(\frac{Qr_{\perp}}{2}\right)\sigma_{q\bar{q}}\left(\frac{1}{2},r_{\perp}\right)\right|^{2}\,.\label{eq:NLOBmu}
\end{equation}
Note that $\frac{d\delta Z}{d\mu}=\frac{d\delta Z_{\gamma}\left(\frac{1}{2},r_{\perp}\right)}{d\mu}$.
This means that the dependence on $\mu$ of the wave-functions of
the photon and heavy quarkonium is the same. This should not be surprising, since in the hard scattering limit both come from a perturbative one-gluon exchange between the quarks.

Now we focus on the dependence on $x_{0}$. For the longitudinal polarization state where the wavefunction is azimuthally symmetric it is convenent to work again with the angularly integrated wave function 
We  first note that 
\begin{equation}
\int\,d\theta_{r}\frac{d}{dx_{0}}\int\,dz\left.\left\{ \Psi_{HQ}^{q\bar{q}}\right\} _{\lambda_{1}\lambda_{2}}^{3}(z,r_{\perp})\right|_{Fig.\ref{fig:kernelr-1}}=-\frac{4\alpha_{s}C_{F}}{x_{0}^{2}}\left[\frac{1}{2}-x_{0}\left(\log\left(\frac{mx_{0}r_{\perp}}{2}\right)+\gamma_{E}+\frac{1}{2}\right)\right]\int\,d\lambda\phi_{q\bar{q}}^{3}(\lambda,{\bf {\bf 0}})\,.
\end{equation}
Using this and computing $\frac{d\delta Z}{dx_{0}}$ we get
\begin{align}
\frac{d}{dx_{0}}\left.\frac{d\sigma_{T,L}^{\gamma^{*}+N\to HQ+N}}{dt}\right|_{NLO,HQ} & =-\frac{\alpha_{s}C_{F}e^{2}e_{f}^{2}Q^{2}N_{c}}{128\pi^{4}x_{0}^{2}}\left|\sum_{h,\bar{h}}\int\frac{\,d\lambda}{4\pi}\phi^{q\bar{q}}\right|^{2}\left|\int\,d^{2}r_{\perp}K_{0}\left(\frac{Qr_{\perp}}{2}\right)\sigma_{q\bar{q}}\left(\frac{1}{2},r_{\perp}\right)\right|^{2}\nonumber \\
 & +\frac{e^{2}e_{f}^{2}Q^{2}N_{c}}{128\pi^{3}}\left|\sum_{h,\bar{h}}\int\frac{\,d\lambda}{4\pi}\phi^{q\bar{q}}\right|^{2}\left(\int\,d^{2}r_{\perp}\frac{d\delta Z_{\gamma}(\frac{1}{2},r_{\perp})}{dx_{0}}K_{0}\left(\frac{Qr_{\perp}}{2}\right)\sigma_{q\bar{q}}\left(\frac{1}{2},r_{\perp}\right)\right) \nonumber \\
 & \quad \quad \times\left(\int\,d^{2}r'_{\perp}K_{0}\left(\frac{Qr'_{\perp}}{2}\right)\sigma_{q\bar{q}}\left(\frac{1}{2},r'_{\perp}\right)\right)^{*}\,.\label{eq:NLOBx}
\end{align}
The first term in the rhs will be cancelled by the dependence of the
non-relativistic quarkonium wave-function on $x_{0}$, c.f. \eq\nr{eq:Coulombsin}.
The other terms have the same dependence on $x_{0}$ as the contribution
of the NLO corrections to the photon wave-function. 

Now we move to the dilute limit, in which we wish to obtain an analytic
expression in the limit $Q\gg m$. Note that this implies that if
$Qr_{\perp}\sim1$ then $mr_{\perp}\ll1$ while if $mr_{\perp}\sim1$
then the photon wave-function is exponentially suppressed. Therefore,
we can use the quarkonium wave-function in the $mr_{\perp}\ll1$ limit.
As a starting point, we take \eq\nr{eq:Cqqtoqqlong} 
\begin{multline}
\left.\int\,d\theta_{r}C_{q\bar{q}\gets q\bar{q}}^{0}(z,{\bf r}_{\perp};\lambda_{1},\lambda_{2};\lambda_{1}',\lambda_{2}')_{long}\right|_{mr_{\perp}\ll1}=8\pi^{2}\delta(z-\frac{1}{2})(1+\delta Z)\delta_{\lambda_{1}\lambda_{1}'}\delta_{\lambda_{2}\lambda_{2}'}
\\
-\frac{4g^{2}C_{F}z(1-z)\delta_{\lambda_{1},\lambda_{1}'}\delta_{\lambda_{2},\lambda_{2}'}}{\left(z-\frac{1}{2}\right)^{2}}
\bigg\{\log\left(\frac{\tau}{2}\right)+\gamma_{E}
\\
+
\left(\theta(z-\frac{1}{2})(1-z)+\theta(\frac{1}{2}-z)z\right)
\left[\frac{\left(z-\frac{1}{2}\right)^{2}-\frac{1}{2}}{z(1-z)}\left(\log\left(\frac{\tau}{2}\right)+\gamma_{E}+\frac{1}{2}\right)+\frac{\left(z-\frac{1}{2}\right)^{2}}{2z(1-z)}\right]
\bigg\}\,.\label{eq:Cqqtoqqlong-1}
\end{multline}
It is convenient to separate this into a $r_{\perp}$ independent and
$r_{\perp}$ dependent piece as
\begin{multline}
\lim_{r_{\perp}\to0}\int\,d\theta_{r}C_{q\bar{q}\gets q\bar{q}}^{0}(z,{\bf r}_{\perp};\lambda_{1},\lambda_{2};\lambda_{1}',\lambda_{2}')_{long}=8\pi^{2}\delta(z-\frac{1}{2})(1+\delta Z)\delta_{\lambda_{1}\lambda_{1}'}\delta_{\lambda_{2}\lambda_{2}'}
+ \delta C(z)
\\
-4g^{2}C_{F}\delta_{\lambda_{1},\lambda_{1}'}\delta_{\lambda_{2},\lambda_{2}'}\log\left(Q\sqrt{z(1-z)}r_{\perp}\right)\left[\theta\left(z-\frac{1}{2}\right)(1-z)\left(1+\frac{1}{z-\frac{1}{2}}\right)+\theta\left(\frac{1}{2}-z\right)z\left(1+\frac{1}{\frac{1}{2}-z}\right)\right]\,,
\end{multline}
where
\begin{multline}
\delta C(z) =-\frac{4g^{2}C_{F}z(1-z)\delta_{\lambda_{1},\lambda_{1}'}\delta_{\lambda_{2},\lambda_{2}'}}{\left(z-\frac{1}{2}\right)^{2}}\left[\log\left(\frac{m\left|z-\frac{1}{2}\right|}{Q\sqrt{z(1-z)}}\right)+\gamma_{E}\right.
\\
  \left.+(\theta(z-\frac{1}{2})(1-z)+\theta(\frac{1}{2}-z)z)\left[\frac{\left(z-\frac{1}{2}\right)^{2}-\frac{1}{2}}{z(1-z)}\left(\log\left(\frac{m\left|z-\frac{1}{2}\right|}{Q\sqrt{z(1-z)}}\right)+\gamma_{E}+\frac{1}{2}\right)+\frac{\left(z-\frac{1}{2}\right)^{2}}{2z(1-z)}\right]\right]\,.
\end{multline}
The final result for the $q\bar{q}$ component of the vector meson wave function that we obtain is 
\begin{multline}
  \left.\frac{d\sigma_{T,L}^{\gamma^{*}+N\to HQ+N}}{dt}\right|_{NLO,HQ,dilute}=
  \\
  -\frac{16C_{F}\alpha_{s}e^{2}e_{f}^{2}\sigma_{0}^{2}Q_{s}^{4}N_{c}}{\pi^{2}Q^{6}}\left[\log x_{0}\left(\frac{1}{D-4}+\frac{1}{2}\log\left(\frac{Q^{2}}{4\pi\mu^{2}}\right)+\frac{\gamma_{E}}{2}-1-\log2\right)\right.
  \\
  \left.+\frac{3}{4}\left(\frac{1}{D-4}+\frac{1}{2}\log\left(\frac{m^{2}}{4\pi\mu^{2}}\right)+\frac{\gamma_{E}}{2}\right)-\frac{1}{2}+\frac{Li_{2}(1)}{2}-Li_{2}\left(\frac{1}{2}\right)+\frac{\pi^{2}}{24}-\frac{1}{2}\log2\left(\log\left(\frac{m}{Q}\right)+1\right)\right]\left|\sum_{h,\bar{h}}\int\frac{\,d\lambda}{4\pi}\phi^{q\bar{q}}\right|^{2}\,.
\end{multline}
Note that the divergences of this part, both in $\mu$ and $x_{0}$,
are the same as in the photon wave function~\nr{eq:A,dilute}.

\subsection{Contribution of the $q\bar{q}g$ Fock state}
\begin{figure}[tbh]
\centerline{
\includegraphics[height=4cm]{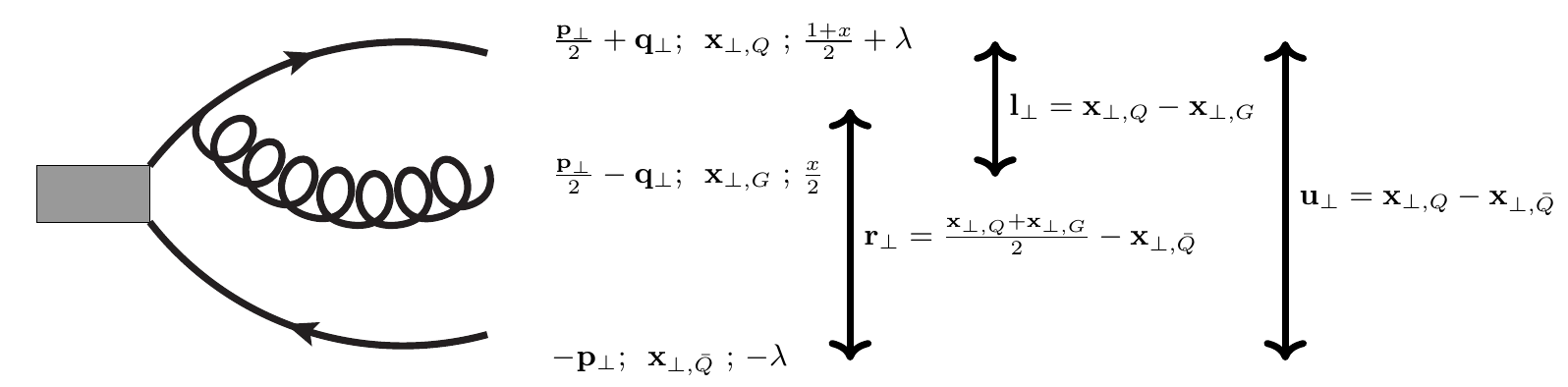}
}
\caption{Transverse momentum, transverse coordinate and  longitudinal momentum fraction assignments for the $q\bar{q}g$ Fock state, and the definitions of the coordinate separations ${\bf r}_{\perp}, {\bf l}_{\perp}, {\bf u}_{\perp}$  in terms of these.\label{fig:qqbgkin}}
\end{figure}

We now move to the $q\bar{q}g$ Fock state, which interacts with the target with the scattering amplitude $\sigma_{q\bar{q}g}$. Note that the NLO correction to the cross section is given by the interference of this NLO correction to the amplitude with the leading order amplitude that only involves a  $q\bar{q}$ state (and of course the complex conjugate). This contribution to the cross section is given by
\begin{equation}
\left.\frac{d\sigma_{T,L}^{\gamma^{*}+N\to HQ+N}}{dt}\right|_{NLO,q\bar{q}g}=\frac{\left|\int\frac{\,d\lambda}{4\pi}\phi^{q\bar{q}}\right|^{2}}{8\pi}\left(\int\,d^{2}r_{\perp}\,d^{2}l_{\perp}\int_{0}^{1}\,\frac{dz}{4\pi}\left(C_{q\bar{q}g\gets q\bar{q}}^{0}\right)^{*}\left(\Psi_{\gamma^{*}}^{q\bar{q}g}\right)_{L}\sigma_{q\bar{q}g}\right)
\left(\int\,d^{2}r'_{\perp}\left(\Psi_{\gamma^{*}}^{q\bar{q}}\right)_{L}\sigma_{q\bar{q}}\right)^{*}\,,
\end{equation}
where $r_{\perp}$ and $l_{\perp}$ are defined as in \eq\nr{eq:hqqbarqg}, with $r_{\perp}$ roughly  interpreted as the large distance separating the nonrelativistic antiquark from the system of a relativistic quark and gluon, which has a small size $l_{\perp}$. These coordinate assignments are demonstrated in Fig.~\ref{fig:qqbgkin}. 
Since we are working in the hard scattering limit $Q\gg M$, the wave-function of the photon can be taken from the massless quark calculation  in \cite{Beuf:2017bpd},
which using our notation and normalization and fixing the antiquark
to be non-relativistic is:
\begin{align}
\Psi_{\gamma^{*}}^{q\bar{q}g}= & \frac{ee_{f}g\sqrt{N_{c}}Q}{m\sqrt{x(1-x)}}\epsilon_{\lambda_{G}}^{A,j*}\bar{u}(\hat{p}_{RQ,}\lambda_{RQ})T^{A}\slashed{\bar{n}}\left[\left(\frac{2-x}{2}\right)\delta^{jm}+\frac{x}{4}[\gamma^{j},\gamma^{m}]\right]v(mv,\lambda_{\bar{Q}})\mathcal{I}^{m}(a)\nonumber \\
 & -\frac{ee_{f}g\sqrt{N_{c}}Q(1-x)}{m\sqrt{x(1-x)}}\epsilon_{\lambda_{G}}^{A,j*}\bar{u}(\hat{p}_{RQ,}\lambda_{RQ})T^{A}\slashed{\bar{n}}\left[\left(\frac{2+x}{2}\right)\gamma^{jm}-\frac{x}{4}[\gamma^{j},\gamma^{m}]\right]v(mv,\lambda_{\bar{Q}})\mathcal{I}^{m}(b)\,,
\end{align}
where
\begin{equation}
\mathcal{I}^{m}(a)=(\mu^{2})^{2-\frac{D}{2}}\int\frac{\,d^{D-2}p_{\perp}}{(2\pi)^{D-2}}\int\frac{\,d^{D-2}k_{\perp}}{(2\pi)^{D-2}}\frac{k_{\perp}^{m}e^{-i{\bf k}_{\perp}{\bf l}_{\perp}}e^{i{\bf p}_{\perp}({\bf r}_{\perp}+\frac{1-2x}{2}{\bf l}_{\perp})}}{\left(p_{\perp}^{2}+\frac{Q^{2}}{4}\right)\left(k_{\perp}^{2}+2x(1-x)\left(p_{\perp}^{2}+\frac{Q^{2}}{4}\right)\right)}\,,
\end{equation}
and 
\begin{equation}
\mathcal{I}^{m}(b)=(\mu^{2})^{2-\frac{D}{2}}\int\frac{\,d^{D-2}p_{\perp}}{(2\pi)^{D-2}}\int\frac{\,d^{D-2}k_{\perp}}{(2\pi)^{D-2}}\frac{k_{\perp}^{m}e^{i{\bf k}_{\perp}({\bf r}_{\perp}-\frac{{\bf l}_{\perp}}{2})}e^{i{\bf p}_{\perp}(\frac{{\bf r}_{\perp}}{1+x}+\frac{1+2x}{2(1+x)}{\bf l}_{\perp})}}{\left(p_{\perp}^{2}+\frac{(1-x^{2})Q^{2}}{4}\right)\left(k_{\perp}^{2}+\frac{2x}{(1-x)(1+x)^{2}}\left(p_{\perp}^{2}+\frac{(1-x^{2})Q^{2}}{4}\right)\right)}\,.
\end{equation}

As in the inclusive DIS case~\cite{Beuf:2017bpd,Hanninen:2017ddy} there are cancellations of divergences between the $q\bar{q}$ and $q\bar{q}g$ states. To make these manifest, it is useful to separate the contribution into two different terms, by adding and subtracting a term that involves a $q\bar{q}$ scattering amplitude
\begin{align}
\left.\frac{d\sigma_{T,L}^{\gamma^{*}+N\to HQ+N}}{dt}\right|_{NLO,q\bar{q}g}= & \frac{\left|\int\frac{\,d\lambda}{4\pi}\phi^{q\bar{q}}\right|^{2}}{8\pi}\left(\int\,d^{2}r_{\perp}\,d^{2}l_{\perp}\int_{0}^{1}\,\frac{dz}{4\pi}\left(C_{q\bar{q}g\gets q\bar{q}}^{0}\right)^{*}\left(\Psi_{\gamma^{*}}^{q\bar{q}g}\right)_{L}\left(\sigma_{q\bar{q}g}-\sigma_{q\bar{q}}\right)\right)
\left(\int\,d^{2}r'_{\perp}\left(\Psi_{\gamma^{*}}^{q\bar{q}}\right)_{L}\sigma_{q\bar{q}}\right)^{*}\nonumber \\
 & +\frac{\left|\int\frac{\,d\lambda}{4\pi}\phi^{q\bar{q}}\right|^{2}}{8\pi}\left(\int\,d^{2}r_{\perp}\,d^{2}l_{\perp}\int_{0}^{1}\,\frac{dz}{4\pi}\left(C_{q\bar{q}g\gets q\bar{q}}^{0}\right)^{*}\left(\Psi_{\gamma^{*}}^{q\bar{q}g}\right)_{L}\sigma_{q\bar{q}}\right)
 \left(\int\,d^{2}r'_{\perp}\left(\Psi_{\gamma^{*}}^{q\bar{q}}\right)_{L}\sigma_{q\bar{q}}\right)^{*}\,.\label{eq:nloc}
\end{align}
Note that the argument of $\sigma_{q\bar{q}}$ is the separation between
the quark and the antiquark, which is ${\bf u}_{\perp}={\bf r}_{\perp}+\frac{{\bf l}_{\perp}}{2}$
for the $q\bar{q}g$ Fock state wave-function, see  Fig.~\ref{fig:qqbgkin}. 

First, we discuss the possible $\mu$ dependence. Note that this is
related with the limit $l_{\perp}\to0$. In this limit $\sigma_{q\bar{q}g}-\sigma_{q\bar{q}}$
also vanishes. Therefore, we only need to focus on the second line
in eq. (\ref{eq:nloc}). For small values of $mxl_{\perp}$, using
the formula in Appendix~\ref{sec:Useful-Fourier-transforms},
\begin{align}
 & C_{q\bar{q}g\gets q\bar{q}}^{0}\sim\nonumber \\
 & -\frac{ig\left(1-\frac{x}{2}\right)l_{\perp}^{i}{\epsilon_{\perp}^{*}}^{A,j}(\lambda_{G})}{2\pi\sqrt{x(1-x)}ml_{\perp}^{2}}\left(\frac{1}{\pi l_{\perp}^{2}\mu^{2}}\right)^{\frac{D-4}{2}}\Gamma\left(\frac{D-2}{2}\right)\sum_{\lambda_{Q}}\bar{u}(\hat{p}_{RQ},\lambda_{RQ})T^{A}\left[\delta^{ij}-\frac{x}{4\left(1-\frac{x}{2}\right)}[\gamma_{\perp}^{i},\gamma_{\perp}^{j}]\right]\slashed{\bar{n}}u(mv,\lambda_{Q})\,.
\end{align}
Regarding $\Psi_{\gamma^{*}}^{q\bar{q}g}$, we can approximate
\begin{equation}
\mathcal{I}^{m}(a)=-\frac{i2^{D-4}l_{\perp}^{m}\mu^{4-D}}{(4\pi)^{\frac{D-2}{2}}l_{\perp}^{D-2}}\Gamma\left(\frac{D-2}{2}\right)\int\frac{\,d^{D-2}p_{\perp}}{(2\pi)^{D-2}}\frac{e^{i{\bf p}_{\perp}({\bf u}_{\perp}-x{\bf l}_{\perp})}}{\left(p_{\perp}^{2}+\frac{Q^{2}}{4}\right)}\,,
\end{equation}
and
\begin{equation}
\mathcal{I}^{m}(b)=0\,.
\end{equation}
Note that this approximation keeps the correct small $l_{\perp}$
behaviour without introducing additional ultraviolet divergences.
Therefore, in this limit, we can write
\begin{multline}
 \left(C_{q\bar{q}g\gets q\bar{q}}^{0}\right)^{*}\left(\Psi_{\gamma^{*}}^{q\bar{q}g}\right)_{L}=
 \\
  =\frac{ee_{f}g^{2}C_{F}\sqrt{N_{c}}Q}{2\pi^{2}l_{\perp}^{2}}\left(\frac{1}{\pi l_{\perp}^{2}\mu^{2}}\right)^{D-4}\left(\Gamma\left(\frac{D-2}{2}\right)\right)^{2}\frac{1}{x}\left(1-x+\frac{x^{2}}{2}\right)\int\frac{\,d^{D-2}p_{\perp}}{(2\pi)^{D-2}}\frac{e^{i{\bf p}_{\perp}({\bf u}_{\perp}-x{\bf l}_{\perp})}}{\left(p_{\perp}^{2}+\frac{Q^{2}}{4}\right)}\,.
\end{multline}
Performing the integration over $l_{\perp}$, we obtain
\begin{multline}
  \mu^{4-D}\int\,d^{D-2}l_{\perp}\left(C_{q\bar{q}g\gets q\bar{q}}^{0}\right)^{*}\left(\Psi_{\gamma^{*}}^{q\bar{q}g}\right)_{L}
  \\
  =\frac{ee_{f}g^{2}C_{F}\sqrt{N_{c}}Q}{2\pi}\left(\frac{x^{2}}{4\pi\mu^{2}}\right)^{\frac{D-4}{2}}\frac{\Gamma\left(\frac{4-D}{2}\right)}{\Gamma(D-3)}\left(\Gamma\left(\frac{D-2}{2}\right)\right)^{2}\frac{1}{x}\left(1-x+\frac{x^{2}}{2}\right)\int\frac{\,d^{D-2}p_{\perp}}{(2\pi)^{D-2}}\frac{p_{\perp}^{D-4}e^{i{\bf p}_{\perp}{\bf u}_{\perp}}}{\left(p_{\perp}^{2}+\frac{Q^{2}}{4}\right)}\,.
\end{multline}
Note that since we performed approximations valid in the limit $l_{\perp}\to0$
we can only claim at the moment to capture the ultraviolet behaviour,
or in other words, the $\mu$ dependence. This is captured by the
following equation
\begin{equation}
\frac{d}{d\mu}\left(\mu^{4-D}\int\frac{\,dx}{4\pi}\int\,d^{D-2}l_{\perp}\left(C_{q\bar{q}g\gets q\bar{q}}^{0}\right)^{*}\left(\Psi_{\gamma^{*}}^{q\bar{q}g}\right)_{L}\right)=-\frac{ee_{f}\alpha_{s}C_{F}\sqrt{N_{c}}Q}{2\pi^{2}\mu}\left(\log x_{0}+\frac{3}{4}\right)K_{0}\left(\frac{r_{\perp}Q}{2}\right)=-\frac{dZ}{d\mu}\Psi_{\gamma}\left(\frac{1}{2},r_{\perp}\right)\,.
\end{equation}
This is required so that  the $\mu$-dependence of the contribution from the $q\bar{q}g$ Fock state crossing the target shock will cancel against the $\mu$-dependence of quark wavefunction renormalization, which is a contribution where only the $q\bar{q}$ dipole crosses the shock wave.

Now we focus on the dependence on $x_{0}$. In the $x\to x_{0}$
limit, we can write 
\begin{equation}
\Psi_{\gamma^{*}}^{q\bar{q}g}=\frac{ee_{f}g\sqrt{2N_{c}}Q\delta_{\lambda_{RQ,}-\lambda_{\bar{Q}}}}{\sqrt{x}}\epsilon_{\lambda_{G}}^{A,j*}\left(\mathcal{I}^{j}(a)-\mathcal{I}^{j}(b)\right)T^{A}\,,
\end{equation}
where
\begin{align}
\mathcal{I}^{j}(a)-\mathcal{I}^{j}(b) & =(\mu^{2})^{2-\frac{D}{2}}\int\frac{\,d^{D-2}k_{\perp}}{(2\pi)^{D-2}}\frac{k_{\perp}^{j}e^{-i{\bf k}_{\perp}{\bf l}_{\perp}}\left(1-e^{i{\bf k}_{\perp}{\bf u}_{\perp}}\right)}{k_{\perp}^{2}}\int\frac{\,d^{D-2}p_{\perp}}{(2\pi)^{D-2}}\frac{e^{i{\bf p}_{\perp}{\bf u}_{\perp}}}{\left(p_{\perp}^{2}+\frac{Q^{2}}{4}\right)}\nonumber \\
 & =-\frac{i\Gamma\left(\frac{D-2}{2}\right)}{2\pi^{\frac{D-2}{2}}}\left(\frac{l_{\perp}^{j}}{l_{\perp}^{D-2}}+\frac{u_{\perp}^{j}-l_{\perp}^{j}}{|u_{\perp}-l_{\perp}|^{D-2}}\right)(\mu^{2})^{2-\frac{D}{2}}\int\frac{\,d^{D-2}p_{\perp}}{(2\pi)^{D-2}}\frac{e^{i{\bf p}_{\perp}{\bf u}_{\perp}}}{\left(p_{\perp}^{2}+\frac{Q^{2}}{4}\right)}\,.
\end{align}
Looking at the original expressions for $\mathcal{I}^{j}(a)$ and
$\mathcal{I}^{j}(b)$, we might wonder if by simplifying the pole
in $k_{\perp}$ we are modifying the $x_{0}$ dependence. However,
note that for small $x_{0}$ the integral is only sensitive to the
exact value of the pole in the infrared. The infrared limit of $\mathcal{I}^{j}(a)$
cancels with that of $\mathcal{I}^{j}(b)$ for $x$ close to $x_{0}$,
and therefore this sensitivity is cancelled out. In the previous formula
this cancellation is given by the factor $\left(1-e^{i{\bf k}_{\perp}{\bf u}_{\perp}}\right)$.

Regarding $C_{q\bar{q}g\gets q\bar{q}}^{0}$, in the $x\to x_{0}$ limit
we can approximate it by 
\begin{equation}
C_{q\bar{q}g\gets q\bar{q}}^{0}\sim-\frac{ig\Gamma\left(\frac{D-2}{2}\right){\epsilon_{\perp}^{*}}^{A,i}(\lambda_{G})T^{A}\delta_{\lambda_{RQ},\lambda_{Q}}l_{\perp}^{i}}{\sqrt{2x}\pi^{\frac{D-2}{2}}l_{\perp}^{D-2}}\,.
\end{equation}
Therefore, we get coefficient for the $q\bar{q}g$ component in the meson wavefunction as
\begin{align*}
 & \left(C_{q\bar{q}g\gets q\bar{q}}^{0}\right)^{*}\left(\Psi_{\gamma^{*}}^{q\bar{q}g}\right)_{L}\\
 & \sim\frac{ee_{f}g^{2}C_{F}\sqrt{N_{c}}Q\delta_{\lambda_{Q},-\lambda_{\bar{Q}}}}{4\pi^{D-2}x}\left(\Gamma\left(\frac{D-2}{2}\right)\right)^{2}\mu^{4-D}\left(\frac{2({\bf u}_{\perp}-{\bf l}_{\perp}){\bf l}_{\perp}}{|{\bf u}_{\perp}-{\bf l}_{\perp}|^{D-2}l_{\perp}^{D-2}}+\frac{1}{l_{\perp}^{2D-6}}+\frac{1}{|{\bf u}_{\perp}-{\bf l}_{\perp}|^{2D-6}}\right)
 \\
 & \quad \quad \times\int\frac{\,d^{D-2}p_{\perp}}{(2\pi)^{D-2}}\frac{e^{i{\bf p}_{\perp}{\bf u}_{\perp}}}{\left(p_{\perp}^{2}+\frac{Q^{2}}{4}\right)}\\
 & =\frac{g^{2}C_{F}}{2\pi^{D-2}x}\left(\Gamma\left(\frac{D-2}{2}\right)\right)^{2}\mu^{4-D}\left(\frac{2({\bf u}_{\perp}-{\bf l}_{\perp}){\bf l}_{\perp}}{|{\bf u}_{\perp}-{\bf l}_{\perp}|^{D-2}l_{\perp}^{D-2}}+\frac{1}{l_{\perp}^{2D-6}}+\frac{1}{|{\bf u}_{\perp}-{\bf l}_{\perp}|^{2D-6}}\right)\Psi_{\gamma}\left(\frac{1}{2},u_{\perp}\right)\,.
\end{align*}

Using these formulae we can compute the dependence on the cutoff $x_0$ separately for the two parts of the decomposition \nr{eq:nloc}
\begin{align}
\frac{d}{dx_{0}}\int_{x_{0}}^{1}\frac{\,dx}{4\pi}\int\,d^{D-2}u_{\perp}\int\,d^{D-2}l_{\perp}\left(C_{q\bar{q}g\gets q\bar{q}}^{0}\right)^{*}\left(\Psi_{\gamma^{*}}^{q\bar{q}g}\right)_{L}(\sigma_{q\bar{q}g}-\sigma_{q\bar{q}})\nonumber \\
+\frac{d}{dx_{0}}\int_{x_{0}}^{1}\frac{\,dx}{4\pi}\int\,d^{D-2}u_{\perp}\int\,d^{D-2}l_{\perp}\left(C_{q\bar{q}g\gets q\bar{q}}^{0}\right)^{*}\left(\Psi_{\gamma^{*}}^{q\bar{q}g}\right)_{L}\sigma_{q\bar{q}}\,.
\end{align}
The first term gives
\begin{multline}
-\frac{\alpha_{s}C_{F}\mu^{4-D}}{2\pi^{D-2}x_{0}}\left(\Gamma\left(\frac{D-2}{2}\right)\right)^{2}\int\,d^{D-2}u_{\perp}\int\,d^{D-2}l_{\perp}\left(\frac{2({\bf u}_{\perp}-{\bf l}_{\perp}){\bf l}_{\perp}}{|{\bf u}_{\perp}-{\bf l}_{\perp}|^{D-2}l_{\perp}^{D-2}}+\frac{1}{l_{\perp}^{2D-6}}+\frac{1}{|{\bf u}_{\perp}-{\bf l}_{\perp}|^{2D-6}}\right)
\\ \times
(\sigma_{q\bar{q}g}-\sigma_{q\bar{q}})\Psi_{\gamma}\left(\frac{1}{2},u_{\perp}\right)\,.
\end{multline}
Since the $l_{\perp}$ integration introduces no divergences (recall that $(\sigma_{q\bar{q}g}-\sigma_{q\bar{q}})\to 0$ when $l_{\perp}\to 0$) we can
take the exact limit $D=4$. Then we get
\begin{equation}
-\frac{\alpha_{s}C_{F}}{2\pi^{2}x_{0}}\int\,d^{2}u_{\perp}\int\,d^{2}l_{\perp}\frac{u_{\perp}^{2}}{|{\bf u}_{\perp}-{\bf l}_{\perp}|^{2}l_{\perp}^{2}}(\sigma_{q\bar{q}g}-\sigma_{q\bar{q}})\Psi_{\gamma}\left(\frac{1}{2},u_{\perp}\right)=-\frac{1}{2}\int\,d^{2}u_{\perp} \frac{d\sigma_{q\bar{q}}}{dx_{0}}\Psi_{\gamma}\left(\frac{1}{2},u_{\perp}\right)\,,
\end{equation}
where in the last equality we have used the B-JIMWLK equation \cite{Ferreiro:2001qy,Iancu:2000hn,Iancu:2001ad,JalilianMarian:1997dw,JalilianMarian:1997gr,JalilianMarian:1997jx,Kovner:2000pt,Weigert:2000gi}.
The other contribution, depending  only on the $q\bar{q}$ amplitude, that was added and subtracted in the decomposition \nr{eq:nloc} becomes
\begin{align}
 & -\frac{\alpha_{s}C_{F}\mu^{4-D}}{2\pi^{D-2}x_{0}}\left(\Gamma\left(\frac{D-2}{2}\right)\right)^{2}\int\,d^{D-2}u_{\perp}\int\,d^{D-2}l_{\perp}\left(\frac{2({\bf u}_{\perp}-{\bf l}_{\perp}){\bf l}_{\perp}}{|{\bf u}_{\perp}-{\bf l}_{\perp}|^{D-2}l_{\perp}^{D-2}}+\frac{1}{l_{\perp}^{2D-6}}+\frac{1}{|{\bf u}_{\perp}-{\bf l}_{\perp}|^{2D-6}}\right)\sigma_{q\bar{q}}\Psi_{\gamma}\left(\frac{1}{2},u_{\perp}\right)\nonumber \\
 & =\frac{2\alpha_{s}C_{F}}{\pi x_{0}}\int\,d^{D-2}r_{\perp}\left(\frac{1}{D-4}-\frac{\gamma_{E}}{2}-\frac{1}{2}\log\left(\pi r_{\perp}^{2}\mu^{2}\right)\right)\sigma_{q\bar{q}}\Psi_{\gamma}\left(\frac{1}{2},r_{\perp}\right)=-\int\,d^{D-2}r_{\perp}\frac{d}{dx_{0}}Z_{\gamma}\left(\frac{1}{2},r_{\perp}\right)\sigma_{q\bar{q}}\Psi_{\gamma}\left(\frac{1}{2},r_{\perp}\right)\,.
\end{align}

Finally, we put all the pieces together. Note that until now we have
only explicitly shown the computation in the case in which the intermediate
quark is relativistic and the antiquark is non-relativistic. However,
the opposite case also gives a contribution which is, due to symmetry
considerations, exactly equal. Taking this into account we get 
\begin{multline}
\frac{d}{d\mu}\left.\frac{d\sigma_{T,L}^{\gamma^{*}+N\to HQ+N}}{dt}\right|_{NLO,q\bar{q}g}  =-\frac{e^{2}e_{f}^{2}Q^{2}N_{c}}{64\pi}\left|\sum_{h,\bar{h}}\int\frac{\,d\lambda}{4\pi}\phi^{q\bar{q}}\right|^{2}\left(\int\,d^{2}r_{\perp}\frac{dZ}{d\mu}K_{0}\left(\frac{Qr_{\perp}}{2}\right)\sigma_{q\bar{q}}\left(\frac{1}{2},r_{\perp}\right)\right)
\\
 \times\left(\int\,d^{2}r'_{\perp}K_{0}\left(\frac{Qr'_{\perp}}{2}\right)\sigma_{q\bar{q}}\left(\frac{1}{2},r'_{\perp}\right)\right)^{*}\,,\label{eq:NLOCmu}
\end{multline}
and
\begin{align}
\frac{d}{dx_{0}}\left.\frac{d\sigma_{T,L}^{\gamma^{*}+N\to HQ+N}}{dt}\right|_{NLO,q\bar{q}g} & =-\frac{e^{2}e_{f}^{2}Q^{2}N_{c}}{128\pi}\left|\sum_{h,\bar{h}}\int\frac{\,d\lambda}{4\pi}\phi^{q\bar{q}}\right|^{2}\left(\int\,d^{2}r_{\perp}K_{0}\left(\frac{Qr_{\perp}}{2}\right)\frac{d\sigma_{q\bar{q}}}{dx_{0}}\left(\frac{1}{2},r_{\perp}\right)\right)
\nonumber \\
 & \quad \quad \times\left(\int\,d^{2}r'_{\perp}K_{0}\left(\frac{Qr'_{\perp}}{2}\right)\sigma_{q\bar{q}}\left(\frac{1}{2},r'_{\perp}\right)\right)^{*}
 \nonumber \\
 & -\frac{e^{2}e_{f}^{2}Q^{2}N_{c}}{64\pi}\left|\sum_{h,\bar{h}}\int\frac{\,d\lambda}{4\pi}\phi^{q\bar{q}}\right|^{2}\left(\int\,d^{2}r_{\perp}K_{0}\left(\frac{Qr_{\perp}}{2}\right)\frac{d}{dx_{0}}Z_{\gamma}\left(\frac{1}{2},r_{\perp}\right)\sigma_{q\bar{q}}\left(\frac{1}{2},r_{\perp}\right)\right)
 \nonumber \\
 & \quad \quad \times\left(\int\,d^{2}r'_{\perp}K_{0}\left(\frac{Qr'_{\perp}}{2}\right)\sigma_{q\bar{q}}\left(\frac{1}{2},r'_{\perp}\right)\right)^{*}\,.\label{eq:NLOCx}
\end{align}

\subsection{Summary}

In this subsection, we check that the sum of all the contributions to the cross section at NLO accuracy has the expected dependence on the cutoff scales  $\mu$ and $x_{0}$. Let us first remind the reader that the leading order cross section is finite and does not depend on the dimensional regularization scale
\begin{equation}
\frac{d}{d\mu}\left.\frac{d\sigma_{T,L}^{\gamma^{*}+N\to HQ+N}}{dt}\right|_{LO}=0\,,\label{eq:LOmu}.
\end{equation}
Combining this with the scale dependence from the photon wave function~\nr{eq:NLOA}, the vector meson wave function \nr{eq:NLOBmu} and the one-loop correction to the wave function~\nr{eq:NLOCmu} we see that the dependence on $\mu$ cancels
\begin{equation}
\frac{d}{d\mu}\left.\frac{d\sigma_{T,L}^{\gamma^{*}+N\to HQ+N}}{dt}\right|_{LO+NLO}=0\,,
\end{equation}

The overall dependence on $x_0$ must also cancel, when the  B-JIMWLK evolution of the target is taken into account. The leading order cross section depends on the cutoff $x_0$ both through the $x_0$-dependence of the dipole cross section, and of the nonrelativistic wavefunction (this resulted from the fact that the decay width must be independent of $x_0$, see \nr{eq:Coulombsin}):
\begin{align}
\frac{d}{dx_{0}}\left.\frac{d\sigma_{T,L}^{\gamma^{*}+N\to HQ+N}}{dt}\right|_{LO} & =\frac{e^{2}e_{f}^{2}Q^{2}N_{c}}{128\pi}\left|\sum_{h,\bar{h}}\int\frac{\,d\lambda}{4\pi}\phi^{q\bar{q}}\right|^{2}\left(\int\,d^{2}r_{\perp}K_{0}\left(\frac{Qr_{\perp}}{2}\right)\frac{d\sigma_{q\bar{q}}}{x_{0}}\left(\frac{1}{2},r_{\perp}\right)\right)\times\nonumber \\
 & \times\left(\int\,d^{2}r'_{\perp}K_{0}\left(\frac{Qr'_{\perp}}{2}\right)\sigma_{q\bar{q}}\left(\frac{1}{2},r'_{\perp}\right)\right)^{*}\nonumber \\
 & +\frac{\alpha_{s}C_{F}e^{2}e_{f}^{2}Q^{2}N_{c}}{128\pi^{4}x_{0}^{2}}\left|\sum_{h,\bar{h}}\int\frac{\,d\lambda}{4\pi}\phi^{q\bar{q}}\right|^{2}\left|\int\,d^{2}r_{\perp}K_{0}\left(\frac{Qr_{\perp}}{2}\right)\sigma_{q\bar{q}}\left(\frac{1}{2},r_{\perp}\right)\right|^{2}\,.\label{eq:LOx}
\end{align}
Both of these $x_0$ dependences of the leading order cross section are proportional to $\as$, and are thus at the same level as the NLO contributions. Relevant NLO contributions are the ones from the photon wave function (i.e. quark wavefunction renormalization), \eq\nr{eq:NLOA}, from the ``vertex correction'' to the $q\bar{q}$ component of the meson light cone wave function in 
\nr{eq:NLOBx}, and from the $q\bar{q}g$ Fock state, \eq\nr{eq:NLOCx}. All of these together are needed to make the cross section at NLO independent of the cutoff, $x_0$, naturally up to terms of higher order in $\as$:
\begin{equation}
\frac{d}{dx_{0}}\left.\frac{d\sigma_{T,L}^{\gamma^{*}+N\to HQ+N}}{dt}\right|_{LO+NLO}=0\,.
\end{equation}
This demonstrates that the leading high energy behavior (i.e. dependence on the cutoff $x_0$) can be factorized into B-JIMWLK evolution of the target. Note that our discussion here has been framed in the language of a simple small-$x$ factorization procedure where the amplitudes $\sigma_{q\bar{q}}$ and $\sigma_{q\bar{q}g}$ depend on the cutoff $x_0$. It is known for other processes (see e.g. \cite{Ducloue:2016shw}) that this can result in an oversubtraction that can give rise to unphysical behavior for the NLO cross section. We expect this problem to require, just as for inclusive hadron production or inclusive DIS \cite{Iancu:2016vyg,Ducloue:2017mpb,Beuf:2017bpd,Ducloue:2017ftk}, taking the amplitude to in fact depend on the gluon longitudinal momentum fraction ($\lambda_G$ in Eq.~(\ref{eq:46})) in the appropriate way. This modification is formally of higher order in $\as$, and does not affect our conclusion that the leading high energy logarithm can be factorized into B-JIMWLK evolution.

\section{Conclusions\label{sec:Conclusions}}

In this manuscript, we have studied the light-cone wave function of
heavy quarkonium in the non-relativistic limit with a focus on future
applications within the dipole model. In section \ref{sec:The-dipole-approach}
we discussed how to combine the dipole model with the non-relativistic
power counting. Within this scope, the main results of this work are
the NLO correction to the $q\bar{q}$ component, given in eqs. (\ref{eq:Cqqtoqqlong})
and (\ref{eq:Cqqtoqqtran}), and the leading contribution to the $q\bar{q}g$
component, given in eq. (\ref{eq:qbqgm}) in momentum space and in
eq. (\ref{eq:hqqbarqg}) in coordinate space.

We have performed several cross-checks of these results:
\begin{itemize}
\item We have checked that the light-cone distribution amplitudes that are
obtained from the light-cone wave functions fulfil ERBL equation.
\item We have confirmed that, for the longitudinal case, we recover the
results of \cite{Barbieri:1975ki} for radiative corrections to S-wave
quarkonium decay.
\item Finally, we have also checked that using the light-cone wave functions
that we found in the longitudinal case to compute exclusive quarkonium
production in the limit $Q\gg m$ we get that all divergences vanish
at NLO assuming B-JIMWLK evolution of the target.
\end{itemize}

The importance of our results lies in a future application to the
computation of quarkonium exclusive production at next-to-leading
order. This would be the heavy quark analogue of the exclusive light meson production  calculation of Ref.~\cite{Boussarie:2016bkq}, which is performed in a slightly different formalism but uses the same physical picture of high energy scattering. The heavy quark mass introduces additional complications on one hand, but simplifies the description of the meson on the other, due to the nonrelativistic nature of the bound state. 
For this the photon light-cone wave function with finite mass
effects is needed, this result will appear in the near future, following the massless case~\cite{Beuf:2017bpd}. Overall, we believe our
work here is a necessary part of a broad effort to increase to next-to-leading order the
accuracy of computations in the dipole picture for a variety of processes. Such improvements are urgently needed to fully confront QCD theory in the saturation regime with experimental data from an active program at the LHC and from a future Electron-Ion Collider.

\begin{acknowledgments}

We are thankful to G. Beuf, V. Guzey and H. Mäntysaari  for discussions. This work has been supported by the Academy of Finland, projects No. 267321, 303756  and No. 321840 and by the European Research Council (ERC) under the European Union’s Horizon 2020 research and innovation programme (grant agreement No ERC-2015-CoG-681707). The work of M.A.E. was supported by the Academy of Finland project 297058, by \textit{Ministerio de Ciencia e Innovacion} of Spain under project FPA2017-83814-P and Maria de Maetzu Unit of Excellence MDM-2016- 0692, by Xunta de Galicia and FEDER. The content of this article does not reflect the official opinion of the European Union and responsibility for the information and views expressed therein lies entirely with the authors.

\end{acknowledgments}
\appendix

\section{Quark wave-function renormalization in light cone gauge}
\label{sec:The-wave-function-renormalizatio}

\subsection{Calculation of renormalization coefficient}
In this appendix, we compute the wave-function renormalization in
light-cone gauge by studying the residue of the quark propagator.
In the literature, this has been obtained before \cite{Mustaki:1990im}.
However, we want to obtain a result that is consistent with our regularization
scheme. The NLO correction to the quark propagator can be written
as 

\begin{equation}
\frac{i(\slashed{p}+m)}{p^{2}-m^{2}+i\epsilon}\Sigma(p)\frac{i(\slashed{p}+m)}{p^{2}-m^{2}+i\epsilon}\,,
\end{equation}
where $\Sigma(p)$ is the quark self-energy
\begin{equation}
\Sigma(p)=-\frac{ig^{2}C_{F}}{4}\int_{x_{0}p^{+}}^{p^{+}}\frac{\,dq^{+}}{2\pi}\frac{1}{q^{+}(p^{+}-q^{+})}\int\frac{\,d^{D-2}q_{\perp}}{(2\pi)^{D-2}}\frac{\gamma^{\mu}(\slashed{\hat{p}}_{Q}+m)\gamma^{\nu}d_{\mu\nu}(\hat{q})}{p^{-}-\hat{q}^{-}-\hat{p}_{Q}^{-}+i\epsilon}\,.
\end{equation}
Due to symmetry considerations, the self-energy can be decomposed
into three different terms
\begin{equation}
\Sigma(p)=A(p^{2})m+B(p^{2})(\slashed{p}-m)+C(p^{2})\slashed{v}\,,
\end{equation}
where $v$ is a vector orthogonal to $p$. Because there is no preferred
direction in the transverse plane we can choose it to be 
\begin{equation}
v=\left(p^{+},-\frac{p_{\perp}^{2}+p^{2}}{2p^{+}},{\bf 0}\right)\,.
\end{equation}
The three components fulfil the following equations
\begin{eqnarray}
A(p^{2})&=&\frac{1}{4}\left(\frac{1}{m}Tr(\Sigma(p))+\frac{1}{p^{2}}Tr(\slashed{p}\Sigma(p))\right)\,,
\\
B(p^{2})&=&\frac{1}{4p^{2}}Tr(\slashed{p}\Sigma(p))\,,
\\
C(p^{2})&=&\frac{1}{4\bar{n}\cdot p}\left(Tr(\slashed{\bar{n}}\Sigma(p))-\frac{\bar{n}\cdot p}{p^{2}}Tr(\slashed{p}\Sigma(p))\right)\,.
\end{eqnarray}

It is important to note that, in light-cone gauge, the good and the
bad components of the quark field can renormalize differently
\begin{equation}
\psi_{0}=\left(\sqrt{Z_{+}}\frac{\slashed{n}\slashed{\bar{n}}}{2}+\sqrt{Z_{-}}\frac{\slashed{\bar{n}}\slashed{n}}{2}\right)\psi\,.
\end{equation}
However, the good component is the only dynamical field and our main
interest. Therefore we are going to identify $Z$ with $Z_{+}$. Taking
this into account we arrive at the conclusion that in light-cone gauge
the procedure to find the wave-function renormalization from the quark
propagator is to compute 
\begin{equation}
\frac{\slashed{n}\slashed{\bar{n}}}{2}\frac{i(\slashed{p}+m)}{p^{2}-m^{2}+i\epsilon}\Sigma(p)\frac{i(\slashed{p}+m)}{p^{2}-m^{2}+i\epsilon}\frac{\slashed{\bar{n}}\slashed{n}}{2}\,,
\end{equation}
and look for the residue of the pole. This procedure gives
\begin{equation}
\delta Z=iB(m^{2})+2im^{2}A'(m^{2})-iC(m^{2})\,,
\end{equation}
where $A'(m^{2})=\left.\frac{dA(p^{2})}{dp^{2}}\right|_{p^{2}=m^{2}}$.

Let us start with the computation of $Tr(\slashed{\bar{n}}\Sigma(p))$
\begin{equation}
Tr(\slashed{\bar{n}}\Sigma(p))=-i(D-2)g^{2}C_{F}\int_{x_{0}p^{+}}^{p^{+}}\frac{\,dq^{+}}{2\pi}\frac{1}{q^{+}}\int\frac{\,d^{D-2}q_{\perp}}{(2\pi)^{D-2}}\frac{1}{p^{-}-\hat{q}^{-}-\hat{p}_{Q}^{-}+i\epsilon}\,,
\end{equation}
the denominator can be rewritten as 
\begin{equation}
p^{-}-\hat{q}^{-}-\hat{p}_{Q}^{-}=\frac{-p^{+}}{2q^{+}(p^{+}-q^{+})}\left(\lambda_{\perp}^{2}+\frac{-q^{+}(p^{+}-q^{+})(p^{2}-m^{2})+m^{2}(q^{+})^{2}}{(p^{+})^{2}}\right)\,,\label{eq:selfedem}
\end{equation}
where
\begin{equation}
{\bf \lambda}_{\perp}={\bf q}_{\perp}+\frac{q^{+}}{p^{+}}{\bf p}_{\perp}\,.
\end{equation}
Using this we obtain
\begin{equation}
Tr(\slashed{\bar{n}}\Sigma(p))=\frac{i(D-2)g^{2}C_{F}p^{+}}{4\pi^{2}}\Gamma\left(\frac{4-D}{2}\right)\left(\frac{m^{2}}{4\pi\mu^{2}}\right)^{\frac{D-4}{2}}\int_{x_{0}}^{1}\,dx(1-x)\left(x^{2}-x(1-x)\frac{p^{2}-m^{2}}{m^{2}}\right)^{\frac{D-4}{2}}\,.\label{eq:109}
\end{equation}
However, we are interested in the expansion around $p^{2}\sim m^{2}$
\begin{equation}
Tr(\slashed{\bar{n}}\Sigma(p))=\frac{i(D-2)g^{2}C_{F}p^{+}}{4\pi^{2}}\Gamma\left(\frac{4-D}{2}\right)\left(\frac{m^{2}}{4\pi\mu^{2}}\right)^{\frac{D-4}{2}}\int_{x_{0}}^{1}\,dx(1-x)x^{D-4}\left(1-\frac{(D-4)(1-x)(p^{2}-m^{2})}{2m^{2}x}\right)+\mathcal{O}((p^{2}-m^{2})^{2})\,.
\end{equation}

Now we focus on $Tr(\Sigma(p))$
\begin{align}
Tr(\Sigma(p))= & img^{2}(D-2)C_{F}\int_{x_{0}p^{+}}^{p^{+}}\frac{\,dq^{+}}{2\pi}\frac{1}{q^{+}(p^{+}-q^{+})}\int\frac{\,d^{D-2}q_{\perp}}{(2\pi)^{D-2}}\frac{1}{p^{-}-\hat{q}^{-}-\hat{p}_{Q}^{-}+i\epsilon}\nonumber \\
 & =-\frac{i(D-2)g^{2}C_{F}m}{4\pi^{2}}\Gamma\left(\frac{4-D}{2}\right)\left(\frac{m^{2}}{4\pi\mu^{2}}\right)^{\frac{D-4}{2}}\int_{x_{0}}^{1}\,dxx^{D-4}\left(1-\frac{(D-4)(1-x)(p^{2}-m^{2})}{2m^{2}x}\right)+\mathcal{O}((p^{2}-m^{2}))\,.
\end{align}
Finally, 
\begin{align}
Tr(\slashed{p}\Sigma(p)) & =ig^{2}C_{F}\int_{x_{0}p^{+}}^{p^{+}}\frac{\,dq^{+}}{2\pi}\frac{1}{q^{+}(p^{+}-q^{+})}\int\frac{\,d^{D-2}q_{\perp}}{(2\pi)^{D-2}}\frac{1}{p^{-}-\hat{q}^{-}-\hat{p}_{Q}^{-}+i\epsilon}\left(\frac{(p^{2}-m^{2})(-2p^{+}+\frac{6-D}{2}q^{+})}{q^{+}}-(D-4)m^{2}\right)\nonumber \\
 & =\frac{ig^{2}C_{F}}{4\pi^{2}}\Gamma\left(\frac{4-D}{2}\right)\left(\frac{m^{2}}{4\pi\mu^{2}}\right)^{\frac{D-4}{2}}\times\nonumber \\
 & \times\int_{0}^{1}\,dxx^{D-4}\left(1-\frac{(D-4)(1-x)(p^{2}-m^{2})}{2m^{2}x}\right)\left(\frac{(p^{2}-m^{2})(2-\frac{6-D}{2}x)}{x}+(D-4)m^{2}\right)+\mathcal{O}((p^{2}-m^{2})^{2})\,.
\end{align}

Using the previous results and expanding around $D\sim4$ we can obtain
the terms that are needed to compute $\delta Z$
\begin{eqnarray}
B(m^{2})&=&-\frac{ig^{2}C_{F}}{8\pi^{2}}
\,,
\\
A'(m^{2})&=&\frac{ig^{2}C_{F}}{8\pi^{2}m^{2}}\left[\left(\frac{1}{D-4}+\frac{1}{2}\log\left(\frac{m^{2}}{4\pi\mu^{2}}\right)+\frac{\gamma_{E}}{2}\right)(2\log x_{0}+1)+\log x_{0}+\frac{1}{2}+(\log x_{0})^{2}\right]\,,
\\
C(m^{2})&=&-\frac{ig^{2}C_{F}}{8\pi^{2}}\left(\frac{1}{D-4}+\frac{1}{2}\log\left(\frac{m^{2}}{4\pi\mu^{2}}\right)+\frac{\gamma_{E}}{2}-2\right)\,.
\end{eqnarray}
Finally, we obtain the wave-function renormalization
\begin{align}
\delta Z & =-\frac{g^{2}C_{F}}{8\pi^{2}}\left[2\left(\frac{1}{D-4}+\frac{1}{2}\log\left(\frac{m^{2}}{4\pi\mu^{2}}\right)+\frac{\gamma_{E}}{2}\right)(2\log x_{0}+1)+2\log x_{0}+2(\log x_{0})^{2}\right.\nonumber \\
 & \left.+\frac{1}{D-4}+\frac{1}{2}\log\left(\frac{m^{2}}{4\pi\mu^{2}}\right)+\frac{\gamma_{E}}{2}-2\right]\,.\label{eq:wfmway}
\end{align}
This result agrees with \cite{Mustaki:1990im} if we assume that they have neglected the combination
\begin{equation}
(4\log x_{0}+3)\frac{\gamma_{E}}{2}+\frac{3}{2}\log\left(\frac{m^{2}}{4\pi\mu^{2}}\right)-2\,,
\end{equation}
probably due to a difference in the renormalization prescription.
Additionally, we have checked that starting from eq. (3.28) in \cite{Mustaki:1990im}
we also obtain our \eq\nr{eq:wfmway}.

\subsection{Non-relativistic subtraction}

Now we study the contribution to the wave-function renormalization
of non-relativistic momenta. The motivation is to compute how $\int_{0}^{1}\,d\lambda\phi(\lambda,{\bf 0})$
depends on the cut-off $x_{0}$. To do this it is useful to understand
how to characterize non-relativistic momenta in light-cone coordinates.
The non-relativistic character is given by the conditions
\begin{equation}
p_{\perp}\ll m\,,
\end{equation}
and
\begin{equation}
p^{+}-\frac{m}{\sqrt{2}}\ll m\,.
\end{equation}
The - component of the momentum, in general, is given by
\begin{equation}
p^{-}=\frac{p_{\perp}^{2}+m^{2}}{2p^{+}}+\frac{p^{2}-m^{2}}{2p^{+}}\,.
\end{equation}
Using the above conditions this can be approximated by
\begin{equation}
p^{-}=\sqrt{2}m-p^{+}+\frac{2\left(p^{+}-\frac{m}{\sqrt{2}}\right)^{2}+p_{\perp}^{2}}{\sqrt{2}m}+\frac{p^{2}-m^{2}}{\sqrt{2}m}+\mathcal{O}\left(\frac{\left(p^{+}-\frac{m}{\sqrt{2}}\right)^{3}}{m^{2}},\frac{p_{\perp}^{3}}{m^{2}}\right)\,.
\end{equation}
Using this approximation we obtain the non-relativistic component
of the self-energy
\begin{equation}
\Sigma_{NR}(p)=\frac{ig^{2}C_{F}}{2\sqrt{2}m}\int_{x_{\Lambda}p^{+}}^{x_{0}p^{+}}\frac{\,dq^{+}}{2\pi}\frac{1}{q^{+}}\int\frac{\,d^{D-2}q_{\perp}}{(2\pi)^{D-2}}\frac{\gamma^{\mu}(\slashed{\hat{p}}_{Q}+m)\gamma^{\nu}d_{\mu\nu}(\hat{q})}{\frac{q_{\perp}^{2}+2(q^{+})^{2}}{2q^{+}}+\frac{p^{2}-m^{2}}{\sqrt{2}m}}\,.
\end{equation}
The infrared cut-off $x_{\Lambda}$ is such that $m\gg x_{0}p^{+}>x_{\Lambda}p^{+}\gg mv$.
It is introduced because we are only interested in the derivative
with respect to $x_{0}$ and because we want to be sure that no resummation
(due to the non-relativistic nature) is required. In other words,
we only care here about the ultraviolet part of the non-relativistic
contribution. Proceeding in a way completely analogous as the computation
in the general case we obtain
\begin{equation}
\delta Z_{NR}=\frac{g^{2}C_{F}}{4\pi^{2}}\log\left(\frac{x_{0}}{x_{\Lambda}}\right)+\frac{g^{2}C_{F}}{2\pi^{2}}\left(\frac{1}{D-4}+\frac{1}{2}\log\left(\frac{m^{2}}{4\pi\mu^{2}}\right)+\frac{\gamma_{E}}{2}\right)\log\left(\frac{x_{0}}{x_{\Lambda}}\right)+\frac{g^{2}C_{F}}{4\pi^{2}}((\log x_{0})^{2}-(\log x_{\Lambda})^{2})\,.
\end{equation}

\section{Relativistic correction to the $q\bar{q}$ component of the light-cone
wave function\label{sec:Relativistic-correction-to}}

Our starting point is the Bethe-Salpeter equation in the momentum-space
form, which we write using a notation similar to the one used in \cite{Bodwin:2006dm}
\begin{equation}
\tilde{\Psi}_{ab}^{BS}(p)=\int\frac{\,d^{4}q}{(2\pi)^{4}}[S_{F}^{Q}(p)]_{aa'}[K(p-q)]_{a'a'',b''b'}\tilde{\Psi}_{a''b''}^{BS}(q)[S_{F}^{\bar{Q}}(p)]_{b'b}\,,\label{eq:BS}
\end{equation}
where 
\begin{equation}
[S_{F}^{Q}(p_{Q})]_{a'a}=\frac{[\slashed{p}_{Q}+m]_{a'a}}{p_{Q}^{2}-m^{2}+i\epsilon}\,,
\end{equation}
and 
\begin{equation}
[S_{F}^{\bar{Q}}(p_{\bar{Q}})]_{bb'}=\frac{[\slashed{p}_{\bar{Q}}-m]_{bb'}}{p_{\bar{Q}}^{2}-m^{2}+i\epsilon}\,.
\end{equation}

The subindices refer to the spinor structure and we do not write explicitly
the colour indices because quarkonium is assumed to be in a singlet
state. Here $K$ is the kernel of the Bethe-Salpeter equation and $\tilde{\Psi}_{ab}^{BS}(p)$
is the Bethe-Salpeter wave function, which depends on the 4-momentum.
Since we are interested in the situation described in fig. \ref{fig:kernelr-1}
we have to take the limit in which $p$ is relativistic and $q$ is
non-relativistic. This limit has two important consequences in eq.
(\ref{eq:BS}):
\begin{enumerate}
\item $K(p-q)$ can be approximated by $K(p)$. Then we end up integrating
the wave-function on the rhs over all possible non-relativistic momenta,
this is nothing but the non-relativistic wave function at the origin. 
\item Since $p$ is a relativistic momentum (meaning that $p^{2}\apprge m^{2}$)
we can safely compute $K(p)$ in perturbation theory, which gives
using our conventions
\begin{equation}
[K(p)]_{a'a'',b''b'}=ig^{2}C_{F}\gamma_{a'a''}^{\mu}\gamma_{b''b'}^{\nu}\frac{d_{\mu\nu}(p)}{p^{2}+i\epsilon}\,.
\end{equation}
\end{enumerate}
Note that eq. (\ref{eq:BS}) is written in terms of 4-momenta which
can be off-shell. To obtain the light-cone wave function first we
rewrite the equation in terms of the on-shell momenta used in light-cone
perturbation theory (which we denote by $\hat{p}$) using the relation
\begin{equation}
p^{\mu}=\hat{p}^{\mu}+\frac{p^{2}-m^{2}}{2\bar{n}p}\bar{n}^{\mu}\,.
\end{equation}
Then we obtain 
\begin{align}
 & \tilde{\Psi}_{ab}^{R}(p)=\frac{ig^{2}C_{F}}{2\sqrt{2}M^{3}z(1-z)(z-\frac{1}{2})}\sum_{\lambda_{1},\lambda_{2}}[\frac{u(\hat{p}_{Q},\lambda_{1})\bar{u}(\hat{p}_{Q},\lambda_{1})}{\left(\frac{M}{2\sqrt{2}}+p^{-}-\hat{p}_{Q}^{-}+i\epsilon\right)}\gamma^{\mu}\tilde{\Psi}^{NR}(x=0)\gamma^{\nu}\frac{v(\hat{p}_{\bar{Q}},\lambda_{2})\bar{v}(\hat{p}_{\bar{Q}},\lambda_{2})}{\left(\frac{M}{2\sqrt{2}}-p^{-}-\hat{p}_{\bar{Q}}^{-}+i\epsilon\right)}]_{ab}\nonumber \\
 & \left(\frac{d_{\mu\nu}(\hat{p})}{p^{-}-\hat{p}^{-}+isgn(z-\frac{1}{2})\epsilon}-\frac{2\sqrt{2}\bar{n}_{\mu}\bar{n}_{\nu}}{M(z-\frac{1}{2})}\right)\,,
\end{align}
where we used the superindex $R$ and $NR$ to denote respectively
the relativistic and non-relativistic component of the wave function
and $M$ is the mass of the quarkonium state. Integrating the previous
formula over $p^{-}$ and using that $M\sim2m$ we obtain \eq\nr{eq:psirel}.

\section{Dirac algebra manipulations\label{sec:Dirac-algebra-manipulations}}

Here we report relations that we used in the main text which involve
the computation of $\bar{u}\slashed{\epsilon}u$ in the case in which
one of the quarks is non-relativistic. The general formulas for any
momenta are well-known in the context of light cone perturbation theory
\cite{Beuf:2017bpd,Lappi:2016oup}. The starting point is
\begin{equation}
\bar{u}(\hat{p}_{RQ},\lambda_{RQ})\slashed{\epsilon}^{*}(\lambda_{G})u(mv,\lambda_{Q})={\epsilon^{-}}^{*}(p_{G},\lambda_{G})\bar{u}(\hat{p}_{RQ},\lambda_{RQ})\slashed{\bar{n}}u(mv,\lambda_{Q})+\bar{u}(\hat{p}_{RQ},\lambda_{RQ})\slashed{\epsilon}_{\perp}^{*}(\lambda_{G})u(mv,\lambda_{Q})\,,
\end{equation}
in which we have used the definition of the light-cone gauge. Now
let us define the good (+) and bad (-) components of the spinor field
\begin{align}
u_{-}(\hat{p},\lambda) & =\frac{\slashed{\bar{n}}\slashed{n}}{2}u(\hat{p},\lambda)\,,
\end{align}
and
\begin{equation}
u_{+}(\hat{p},\lambda)=\frac{\slashed{n}\slashed{\bar{n}}}{2}u(\hat{p},\lambda)\,.
\end{equation}
In terms of these components we get
\begin{multline}
\bar{u}(\hat{p}_{RQ},\lambda_{RQ})\slashed{\epsilon}^{*}(\lambda_{G})u(mv,\lambda_{Q})=  {\epsilon^{-}}^{*}(p_{G},\lambda_{G})\bar{u}_{+}(\hat{p}_{RQ},\lambda_{RQ})\slashed{\bar{n}}u_{+}(mv,\lambda_{Q})+\bar{u}_{+}(\hat{p}_{RQ},\lambda_{RQ})\slashed{\epsilon}_{\perp}^{*}(\lambda_{G})u_{-}(mv,\lambda_{Q})
\\
  +\bar{u}_{-}(\hat{p}_{RQ},\lambda_{RQ})\slashed{\epsilon}_{\perp}^{*}(\lambda_{G})u_{+}(mv,\lambda_{Q})\,.
\end{multline}
Now we use the Dirac equation to obtain
\begin{equation}
u_{-}(mv,\lambda_{Q})=\frac{\slashed{\bar{n}}}{\sqrt{2}}u_{+}(mv,\lambda_{Q})\,,
\end{equation}
and 
\begin{equation}
\bar{u}_{-}(\hat{p}_{RQ},\lambda_{RQ})=\bar{u}_{+}(\hat{p}_{RQ},\lambda_{RQ})(m+\slashed{q}_{\perp})\frac{\slashed{\bar{n}}}{2\sqrt{2}mz_{RQ}}\,.
\end{equation}
Using this we get
\begin{align}
\bar{u}(\hat{p}_{RQ},\lambda_{RQ})\slashed{\epsilon}^{*}(\lambda_{G})u(mv,\lambda_{Q})= & {\epsilon^{-}}^{*}(p_{G},\lambda_{G})\bar{u}_{+}(\hat{p}_{RQ},\lambda_{RQ})\slashed{\bar{n}}u_{+}(mv,\lambda_{Q})+\frac{1}{\sqrt{2}}\bar{u}_{+}(\hat{p}_{RQ},\lambda_{RQ})\slashed{\epsilon}_{\perp}^{*}(\lambda_{G})\slashed{\bar{n}}u_{+}(mv,\lambda_{Q})\nonumber \\
 & +\frac{1}{2\sqrt{2}z_{RQ}}\bar{u}_{+}(\hat{p}_{RQ},\lambda_{RQ})\slashed{\bar{n}}\slashed{\epsilon}_{\perp}^{*}(\lambda_{G})u_{+}(mv,\lambda_{Q})\nonumber \\
 & +\frac{1}{2\sqrt{2}mz_{RQ}}\bar{u}_{+}(\hat{p}_{RQ},\lambda_{RQ})\slashed{q}_{\perp}\slashed{\bar{n}}\slashed{\epsilon}_{\perp}^{*}(\lambda_{G})u_{+}(mv,\lambda_{Q})\,.
\end{align}
Now using the property $p^{\mu}\epsilon_{\mu}(p)=0$ we obtain 
\begin{align}
\bar{u}(\hat{p}_{RQ},\lambda_{RQ})\slashed{\epsilon}^{*}(\lambda_{G})u(mv,\lambda_{Q})= & \frac{({\bf q}_{\perp}{\bf \epsilon}_{\perp}^{*}(\lambda_{G}))(z_{G}+2z_{RQ})}{2\sqrt{2}mz_{G}z_{RQ}}\bar{u}_{+}(\hat{p}_{RQ},\lambda_{RQ})\slashed{\bar{n}}u_{+}(mv,\lambda_{Q})\nonumber \\
 & +\frac{2z_{RQ}-1}{2\sqrt{2}z_{RQ}}\bar{u}_{+}(\hat{p}_{RQ},\lambda_{RQ})\slashed{\epsilon}_{\perp}^{*}(\lambda_{G})\slashed{\bar{n}}u_{+}(mv,\lambda_{Q})\nonumber \\
 & -\frac{1}{4\sqrt{2}mz_{RQ}}\bar{u}_{+}(\hat{p}_{RQ},\lambda_{RQ})\slashed{\bar{n}}[\slashed{q}_{\perp},\slashed{\epsilon}_{\perp}^{*}(\lambda_{G})]u_{+}(mv,\lambda_{Q})\,.
\end{align}
Note that $\slashed{\bar{n}}u=\slashed{\bar{n}}u_{+}$and $\slashed{\bar{n}}u_{-}=0$.
Therefore, we can safely remove the subindex + in the rhs.

\section{Useful Fourier transforms\label{sec:Useful-Fourier-transforms}}

In this appendix, we report several equations which are useful to
transform our results from position to momentum space. Some of them
are taken from \cite{Groote:2018rpb}
\begin{equation}
D^{(\kappa)}(r_{\perp},m)=\int\frac{\,d^{D-2}p_{\perp}}{(2\pi)^{D-2}}\frac{e^{i{\bf p}_{\perp}{\bf r}_{\perp}}}{(p_{\perp}^{2}+m^{2})^{\kappa+1}}=\frac{(m/r_{\perp})^{\frac{D-4-2\kappa}{2}}}{(2\pi)^{\frac{D-2}{2}}2^{\kappa}\Gamma(\kappa+1)}K_{\frac{D-4-2\kappa}{2}}(mr_{\perp})\,.\label{eq:Bessel1}
\end{equation}
The previous formula can be used to derive
\begin{equation}
I_{1}^{(\kappa)}(r_{\perp},m)=\int\frac{\,d^{D-2}p_{\perp}}{(2\pi)^{D-2}}\frac{e^{i{\bf p}_{\perp}{\bf r}_{\perp}}({\bf p}_{\perp}{\bf r}_{\perp})}{(p_{\perp}^{2}+m^{2})^{\kappa+1}}=\frac{ir_{\perp}^{2}(m/r_{\perp})^{\frac{D-2-2\kappa}{2}}}{(2\pi)^{\frac{D-2}{2}}2^{\kappa}\Gamma(\kappa+1)}K_{\frac{D-2-2\kappa}{2}}(mr_{\perp})\,,\label{eq:Bessel2}
\end{equation}
and
\begin{equation}
I_{2}^{(\kappa)}(r_{\perp},m)=\int\frac{\,d^{D-2}p_{\perp}}{(2\pi)^{D-2}}\frac{e^{i{\bf p}_{\perp}{\bf r}_{\perp}}({\bf p}_{\perp}{\bf r}_{\perp})^{2}}{(p_{\perp}^{2}+m^{2})^{\kappa+1}}=\frac{(m/r_{\perp})^{\frac{D-4-2\kappa}{2}}mr_{\perp}}{(2\pi)^{\frac{D-2}{2}}2^{\kappa}\Gamma(\kappa+1)}\left(K_{\frac{D-2-2\kappa}{2}}(mr_{\perp})-mxK_{\frac{D-2\kappa}{2}}(mr_{\perp})\right)\,.
\end{equation}
At several points we also need the expressions for $K_{\frac{D-2}{2}}(mr_{\perp})$
in the limit $mr_{\perp}\ll1$. This can be obtained using eq. (\ref{eq:Bessel2})
and the technique of integration by regions \cite{Beneke:1997zp}
\begin{equation}
K_{\frac{D-2}{2}}(mr_{\perp})\sim\frac{1}{mr_{\perp}}\left(\frac{2}{mr_{\perp}}\right)^{\frac{D-4}{2}}\Gamma\left(\frac{D-2}{2}\right)\,.
\end{equation}

\section{Intermediate results for the evaluation of Fig.~\ref{fig:kernelr-1}}
\label{sec:fig6_intermediate}

Fourier-transforming the contribution in \eq\nr{eq:psirel} from momentum to coordinate space (see Appendix~\ref{sec:Useful-Fourier-transforms}) we can write the contribution as
\begin{align}
\left.\left\{ \Psi_{HQ}^{q\bar{q}}\right\} _{\lambda_{1}\lambda_{2}}^{i}(z,r_{\perp})\right|_{Fig.\ref{fig:kernelr-1}} & =\frac{g^{2}C_{F}\sqrt{z(1-z)}}{2\pi m^{2}\left(z-\frac{1}{2}\right)^{2}}\left[K_{0}(\tau)-\frac{\left(z+\frac{1}{2}\right)\left(\frac{3}{2}-z\right)}{2z(1-z)}(\theta(z-\frac{1}{2})(1-z)+\theta(\frac{1}{2}-z)z)\left(K_{0}(\tau)-\frac{\tau}{2}K_{1}(\tau)\right)\right]\times\nonumber \\
 & \times\sum_{\lambda_{1}',\lambda_{2}'}[\bar{u}(\hat{p}_{Q},\lambda_{1})\slashed{\bar{n}}u(mv,\lambda_{1}')]\int\frac{\,d\lambda}{4\pi}\phi_{q\bar{q}}^{i}(\lambda,{\bf {\bf 0}})[\bar{v}(mv,\lambda_{2}')\slashed{\bar{n}}v(\hat{p}_{\bar{Q}},\lambda_{2})]\nonumber \\
 & +\frac{g^{2}C_{F}}{32\pi m^{2}\sqrt{z(1-z)}}(\theta(z-\frac{1}{2})(1-z)+\theta(\frac{1}{2}-z)z)K_{0}(\tau)\times\nonumber \\
 & \times\sum_{\lambda_{1}',\lambda_{2}'}[\bar{u}(\hat{p}_{Q},\lambda_{1})\slashed{\bar{n}}[\gamma_{\perp}^{i},\gamma_{\perp}^{j}]u(mv,\lambda_{1}')]\int\frac{\,d\lambda}{4\pi}\phi_{q\bar{q}}^{i}(\lambda,{\bf {\bf 0}})[\bar{v}(mv,\lambda_{2}')[\gamma_{\perp}^{j},\gamma_{\perp}^{i}]\slashed{\bar{n}}v(\hat{p}_{\bar{Q}},\lambda_{2})]\nonumber \\
 & -\frac{g^{2}C_{F}\left(z-\frac{1}{2}\right)}{16\pi mr_{\perp}\sqrt{z(1-z)}}(\theta(z-\frac{1}{2})(1-z)+\theta(\frac{1}{2}-z)z)K_{1}(\tau)\times\nonumber \\
 & \times\sum_{\lambda_{1}',\lambda_{2}'}[\bar{u}(\hat{p}_{Q},\lambda_{1})\slashed{\bar{n}}[\slashed{r}_{\perp},\gamma_{\perp}^{i}]u(mv,\lambda_{1}')]\int\frac{\,d\lambda}{4\pi}\phi_{q\bar{q}}^{i}(\lambda,{\bf {\bf 0}})[\bar{v}(mv,\lambda_{2}')[\gamma_{\perp}^{i},\slashed{r}_{\perp}]\slashed{\bar{n}}v(\hat{p}_{\bar{Q}},\lambda_{2})]\nonumber \\
 & +\frac{ig^{2}C_{F}}{4\pi m\sqrt{z(1-z)}}(\theta(z-\frac{1}{2})(1-z)+\theta(\frac{1}{2}-z)z)K_{0}(\tau)\times\nonumber \\
 & \times\sum_{\lambda_{1}',\lambda_{2}'}\left[\left(z+\frac{1}{2}\right)[\bar{u}(\hat{p}_{Q},\lambda_{1})\slashed{\bar{n}}u(mv,\lambda_{1}')]\int\frac{\,d\lambda}{4\pi}\phi_{q\bar{q}}^{i}(\lambda,{\bf {\bf 0}})[\bar{v}(mv,\lambda_{2}')\slashed{r}_{\perp}\slashed{\bar{n}}v(\hat{p}_{\bar{Q}},\lambda_{2})]\right.\nonumber \\
 & \left.+\frac{\left(z-\frac{1}{2}\right)}{2}[\bar{u}(\hat{p}_{Q},\lambda_{1})\slashed{\bar{n}}[\slashed{r}_{\perp},\gamma_{\perp}^{i}]u(mv,\lambda_{1}')]\int\frac{\,d\lambda}{4\pi}\phi_{q\bar{q}}^{i}(\lambda,{\bf {\bf 0}})[\bar{v}(mv,\lambda_{2}')\gamma_{\perp}^{i}\slashed{\bar{n}}v(\hat{p}_{\bar{Q}},\lambda_{2})]\right]\nonumber \\
 & -\frac{ig^{2}C_{F}}{4\pi m\sqrt{z(1-z)}}(\theta(z-\frac{1}{2})(1-z)+\theta(\frac{1}{2}-z)z)K_{0}(\tau)\times\nonumber \\
 & \times\sum_{\lambda_{1}',\lambda_{2}'}\left[\left(\frac{3}{2}-z\right)[\bar{u}(\hat{p}_{Q},\lambda_{1})\slashed{r}_{\perp}\slashed{\bar{n}}u(mv,\lambda_{1}')]\int\frac{\,d\lambda}{4\pi}\phi_{q\bar{q}}^{i}(\lambda,{\bf {\bf 0}})[\bar{v}(mv,\lambda_{2}')\slashed{\bar{n}}v(\hat{p}_{\bar{Q}},\lambda_{2})]\right.\nonumber \\
 & \left.-\frac{\left(z-\frac{1}{2}\right)}{2}[\bar{u}(\hat{p}_{Q},\lambda_{1})\gamma_{\perp}^{i}\slashed{\bar{n}}u(mv,\lambda_{1}')]\int\frac{\,d\lambda}{4\pi}\phi_{q\bar{q}}^{i}(\lambda,{\bf {\bf 0}})[\bar{v}(mv,\lambda_{2}')\slashed{\bar{n}}[\gamma_{\perp}^{i},\slashed{r}_{\perp}]v(\hat{p}_{\bar{Q}},\lambda_{2})]\right]\nonumber \\
 & +\frac{g^{2}C_{F}\left(z-\frac{1}{2}\right)r_{\perp}}{4\pi m\sqrt{z(1-z)}}(\theta(z-\frac{1}{2})(1-z)+\theta(\frac{1}{2}-z)z)K_{1}(\tau)\times\nonumber \\
 & \times\sum_{\lambda_{1}',\lambda_{2}'}[\bar{u}(\hat{p}_{Q},\lambda_{1})\gamma_{\perp}^{i}\slashed{\bar{n}}u(mv,\lambda_{1}')]\int\frac{\,d\lambda}{4\pi}\phi_{q\bar{q}}^{i}(\lambda,{\bf {\bf 0}})[\bar{v}(mv,\lambda_{2}')\gamma_{\perp}^{i}\slashed{\bar{n}}v(\hat{p}_{\bar{Q}},\lambda_{2})]\,,
\label{eq:longwf_v1}
 \end{align}
where $\tau=2m\left(z-\frac{1}{2}\right)r_{\perp}$.

Using the Dirac equation satisfied by the spinors and the tabulated spinor matrix elements 
we can  express the matrix elements \nr{eq:ubarepslashu} and  \nr{eq:vbarepslashv} in terms of the helicities of the quark and antiquark as
\begin{multline}
\bar{u}(\hat{p}_{Q},\lambda_{1})\slashed{\epsilon}u(mv,\lambda_{1}')=\frac{p_{\perp}^{i}\epsilon_{\perp}^{j}}{\sqrt{2z}\left(z-\frac{1}{2}\right)}\left[\left(z+\frac{1}{2}\right)\delta^{ij}-i\left(z-\frac{1}{2}\right)(-1)^{\frac{1-\lambda_1}{2}}\epsilon^{ij}\right]\delta_{\lambda_1,\lambda_1'}\\
-\frac{\sqrt{2}m\left(z-\frac{1}{2}\right)(-1)^{\frac{1-\lambda_1}{2}}\delta_{\lambda_1,-\lambda_1'}}{\sqrt{z}}\epsilon_\perp^i(\delta^{i1}-i(-1)^{\frac{1-\lambda_1}{2}}\delta^{i2}),
\label{eq:ubarepslashu_v1}
\end{multline}
\begin{multline}
\bar{v}(mv,\lambda_{2}')\slashed{\epsilon}v(\hat{p}_{\bar{Q}},\lambda_{2})  =\frac{p_{\perp}^{i}\epsilon_{\perp}^{j}}{\sqrt{2(1-z)}\left(z-\frac{1}{2}\right)}\left[\left(\frac{3}{2}-z\right)\delta^{ij}+i\left(z-\frac{1}{2}\right)(-1)^{\frac{1-\lambda_2}{2}}\epsilon^{ij}\right]\delta_{\lambda_2,\lambda_2'}\\
+\frac{\sqrt{2}m\left(z-\frac{1}{2}\right)(-1)^{\frac{1-\lambda_2}{2}}\delta_{\lambda_2,-\lambda_2'}}{\sqrt{1-z}}\epsilon_\perp^i(\delta^{i1}-i(-1)^{\frac{1-\lambda_2}{2}}\delta^{i2}) .
\label{eq:vbarepslashv_v1}
\end{multline}
Note that here the polarization vector $\epsilon_\perp$ is that of the transversely polarized gluon exchanged in Fig.~\ref{fig:kernelr-1}; the longitudinal polarization for the exchanged gluon is treated as a separate term in \eq\nr{eq:psirel}.

Using the matrix elements such as \nr{eq:ubarepslashu_v1} and \nr{eq:vbarepslashv_v1} in 
\nr{eq:longwf_v1} we can write it in the form
\begin{align}
\left.\left\{ \Psi_{HQ}^{q\bar{q}}\right\} _{\lambda_{1}\lambda_{2}}^{i}(z,r_{\perp})\right|_{Fig.\ref{fig:kernelr-1}} & =\frac{2g^{2}C_{F}z(1-z)}{\pi \left(z-\frac{1}{2}\right)^{2}}\left[K_{0}(\tau)-\frac{\left(z+\frac{1}{2}\right)\left(\frac{3}{2}-z\right)}{2z(1-z)}(\theta(z-\frac{1}{2})(1-z)+\theta(\frac{1}{2}-z)z)\left(K_{0}(\tau)-\frac{\tau}{2}K_{1}(\tau)\right)\right]\times\nonumber \\
 & \times\int\frac{\,d\lambda}{4\pi}\left\{\phi_{q\bar{q}}^{i}(\lambda,{\bf {\bf 0}})\right\}_{\lambda_1,\lambda_2}\nonumber \\
 & -\frac{g^{2}C_{F}}{\pi}(\theta(z-\frac{1}{2})(1-z)+\theta(\frac{1}{2}-z)z)K_{0}(\tau)(-1)^{\frac{2-\lambda_1-\lambda_2}{2}}\left\{\int\frac{\,d\lambda}{4\pi}\phi_{q\bar{q}}^{i}(\lambda,{\bf {\bf 0}})\right\}_{\lambda_1,\lambda_2}\nonumber \\
 & +\frac{g^{2}C_{F}\left(z-\frac{1}{2}\right)mr_\perp}{\pi}(\theta(z-\frac{1}{2})(1-z)+\theta(\frac{1}{2}-z)z)K_{1}(\tau)\times(-1)^{\frac{2-\lambda_1-\lambda_2}{2}}\left\{\int\frac{\,d\lambda}{4\pi}\phi_{q\bar{q}}^{i}(\lambda,{\bf {\bf 0}})\right\}_{\lambda_1,\lambda_2}\nonumber \\
 & +\frac{ig^{2}C_{F}m}{4\pi }(\theta(z-\frac{1}{2})(1-z)+\theta(\frac{1}{2}-z)z)K_{0}(\tau)\times\nonumber \\
 & \times\left[-4\left(z+\frac{1}{2}\right)\left\{\int\frac{\,d\lambda}{4\pi}\phi_{q\bar{q}}^{i}(\lambda,{\bf {\bf 0}})\right\}_{\lambda_1,-\lambda_2}r_{\perp}^{j}((-1)^{\frac{1-\lambda_2}{2}}\delta^{j1}-i\delta^{j2})\right.\nonumber \\
 & \left.+8\frac{\left(z-\frac{1}{2}\right)}{2}(-1)^{\frac{2-\lambda_1-\lambda_2}{2}}\left\{\int\frac{\,d\lambda}{4\pi}\phi_{q\bar{q}}^{i}(\lambda,{\bf {\bf 0}})\right\}_{\lambda_1,-\lambda_2}r_{\perp}^{j}((-1)^{\frac{1-\lambda_2}{2}}\delta^{j1}-i\delta^{j2})\right]\nonumber \\
 & -\frac{ig^{2}C_{F}m}{4\pi }(\theta(z-\frac{1}{2})(1-z)+\theta(\frac{1}{2}-z)z)K_{0}(\tau)\times\nonumber \\
 & \times\left[-4\left(\frac{3}{2}-z\right)\left\{\int\frac{\,d\lambda}{4\pi}\phi_{q\bar{q}}^{i}(\lambda,{\bf {\bf 0}})\right\}_{-\lambda_1,\lambda_2}r_\perp^j((-1)^{\frac{1-\lambda_1}{2}}\delta^{j1}-i\delta^{j2})\right.\nonumber \\
 & \left.-8\frac{\left(z-\frac{1}{2}\right)}{2}(-1)^{\frac{2-\lambda_1-\lambda_2}{2}}\left\{\int\frac{\,d\lambda}{4\pi}\phi_{q\bar{q}}^{i}(\lambda,{\bf {\bf 0}})\right\}_{-\lambda_1,\lambda_2}r_\perp^j((-1)^{\frac{1-\lambda_1}{2}}\delta^{j1}-i\delta^{j2})\right]\nonumber \\
 & +\frac{g^{2}C_{F}\left(z-\frac{1}{2}\right)mr_{\perp}}{\pi}(\theta(z-\frac{1}{2})(1-z)+\theta(\frac{1}{2}-z)z)K_{1}(\tau)\times\nonumber\\
 &\times\left((-1)^{\frac{2-\lambda_1-\lambda_2}{2}}-1\right)\left\{\int\frac{\,d\lambda}{4\pi}\phi_{q\bar{q}}^{i}(\lambda,{\bf {\bf 0}})\right\}_{-\lambda_1,-\lambda_2}\,,
\end{align}
which is further simplified to get the result \nr{eq:longwf}.

\bibliographystyle{JHEP-2modlong}
\bibliography{nrdipole}

\end{document}